\documentstyle[12pt]{article}
\begin{document}
\begin{titlepage}
\begin{flushright}
NTUA--08/01 \\ hep-ph/0111125
\end{flushright}
\vspace{1cm}

\begin{centering}
\vspace{.4in} {\Large {\bf Dimensional Reduction over Coset
Spaces\\ and \\ \vspace{.3cm}Supersymmetry Breaking.}}\\
\vspace{1.5cm}

{\bf P.~Manousselis}$^{a}$ and {\bf G.~Zoupanos}$^{b}$\\
\vspace{.2in} Physics Department, National Technical University,
\\ Zografou
Campus, 157 80 Athens, Greece.\\

\vspace{1.0in}

{\bf Abstract}\\

\vspace{.1in} We address the question of supersymmetry breaking of
a higher dimensional supersymmetric theory due to coset space
dimensional reduction. In particular we study a ten-dimensional
supersymmetric $E_{8}$ gauge theory which is reduced over all
six-dimensional coset spaces. We find that the original
supersymmetry is completely broken in the process of dimensional
reduction when the coset spaces are symmetric. On the contrary
softly broken four-dimensional supersymmetric theories result
when the coset spaces are non-symmetric. From our analysis two
promising cases are emerging which lead to interesting GUTs with
three fermion families in four dimensions, one being
non-supersymmetric and the other softly broken supersymmetric.
\end{centering}
\vspace{3.5cm}

\begin{flushleft}
$^{a}$e-mail address: pman@central.ntua.gr. Supported by
$\Gamma\Gamma$ET  grand 97E$\Lambda$/71.
\\ $^{b}$e-mail address:
George.Zoupanos@cern.ch. Partially supported by EU under the RTN
contract HPRN-CT-2000-00148 and the A.v.Humboldt Foundation.
\end{flushleft}
\end{titlepage}
\section{Introduction}
The celebrated Standard Model (SM) of Elementary Particle Physics
had so far outstanding successes in all its confrontations with
experimental results. However the apparent success of the SM is
spoiled by the presence of a plethora of free parameters mostly
related to the ad-hoc introduction of the Higgs and Yukawa sectors
in the theory. It is worth recalling that the Coset Space
Dimensional Reduction (CSDR) \cite{Manton,Review,Kuby} was
suggesting from the beginning that a unification of the gauge and
Higgs sectors can be achieved in higher dimensions. The
four-dimensional gauge and Higgs fields are simply the surviving
components of the gauge fields of a pure gauge theory defined in
higher dimensions. In the next step of development of the CSDR
scheme, fermions were introduced \cite{Slansky} and then the
four-dimensional Yukawa and gauge interactions of fermions found
also a unified description in the gauge interactions of the higher
dimensional theory. The last step in this unified description in
high dimensions is to relate the gauge and fermion fields that
have been introduced. A simple way to achieve that is to demand
that the higher dimensional gauge theory is\ ${\cal N}= 1$
supersymmetric which requires that the gauge and fermion fields
are members of the same supermultiplet. An additional strong
argument towards higher dimensional supersymmetry including
gravity comes from the stability of the corresponding
compactifying solutions that lead to the four-dimensional theory.

In the spirit described above a very welcome additional input is
that string theory suggests furthermore the dimension and the
gauge group of the higher dimensional supersymmetric theory
\cite{Theisen}. Further support to this unified description comes
from the fact that the reduction of the theory over coset
\cite{Review} and CY spaces \cite{Theisen} provides the
four-dimensional theory with scalars belonging in the fundamental
representation of the gauge group as are introduced in the SM. In
addition the fact that the SM is a chiral theory lead us to
consider $D$-dimensional supersymmetric gauge theories with
$D=4n+2$ \cite{Chapline,Review}, which include the ten dimensions
suggested by the heterotic string theory \cite{Theisen}.

Concerning supersymmetry, the nature of the four-dimensional
theory depends on the corresponding nature of the compact space
used to reduce the higher dimensional theory. Specifically the
reduction over CY spaces leads to supersymmetric theories
\cite{Theisen} in four dimensions, the reduction over symmetric
coset spaces leads to non-supersymmetric theories, while a
reduction over non-symmetric ones leads to softly broken
supersymmetric theories \cite{Pman}. Concerning the latter as
candidate four-dimensional theories that describe the nature, in
addition to the usual arguments related to the hierarchy problem
\cite{Dim}, we should remind a further evidence established in
their favor the last years. It was found that the search for
renormalization group invariant (RGI) relations among parameters
of softly broken supersymmetric GUTs considered as a unification
scheme at the quantum level, could lead to successful predictions
in low energies. More specifically the search for RGI relations
was concerning the parameters of softly broken GUTs beyond the
unification point and could lead even to all-loop finiteness
\cite{Mondragon1,Mondragon2}. On the other hand in the low
energies lead to successful predictions not only for the gauge
couplings but also for the top quark mass, among others, and to
interesting testable predictions for the Higgs mass
\cite{Kobayashi}.

The paper is organized as follows. In section 2 we present various
details of the Coset Space Geometry with emphasis in the inclusion
of torsion and more than one radii when possible. The CSDR scheme
is also presented in sufficient detail to make the paper
self-contained. In section 3 supersymmetry breaking via CSDR is
examined. In 3.1 the issue of supersymmetry breaking in CSDR over
symmetric coset spaces is analyzed with presentation of explicit
examples of reduction of a supersymmetric ten-dimensional $E_{8}$
gauge theory over all symmetric six-dimensional coset spaces.
Section 3.2 contains an explicit and detailed study of the
supersymmetry breaking of the same supersymmetric ten-dimensional
$E_{8}$ gauge theory, compactified over all non-symmetric coset
spaces i.e. $G_{2}/SU(3)$, $Sp(4)/(SU(2) \times U(1))_{non-max.}$
and $SU(3)/U(1) \times U(1)$. In section 4 we present our
conclusions. The appendix A contains the commutation relations on
which the calculations of section 3 are based on, while the
appendix B contains details related to the calculation of the
gaugino mass in the reduction over the $SU(3)/U(1) \times U(1)$
coset space.
\section{The Coset Space Dimensional Reduction.}
Given a gauge theory defined in higher dimensions the obvious way
to dimensionally reduce it is to demand that the field dependence
on the extra coordinates is such that the Lagrangian is
independent of them. A crude way to fulfill this requirement is to
discard the field dependence on the extra coordinates, while an
elegant one is to allow for a non-trivial dependence on them, but
impose the condition that a symmetry transformation by an element
of the isometry group $S$ of the space formed by the extra
dimensions $B$ corresponds to a gauge transformation. Then the
Lagrangian will be independent of the extra coordinates just
because it is gauge invariant. This is the basis of the CSDR
scheme \cite{Manton,Review,Kuby}, which assumes that $B$ is a
compact coset space, $S/R$.

In the CSDR scheme one starts with a Yang-Mills-Dirac Lagrangian,
with gauge group $G$, defined on a
 $D$-dimensional spacetime $M^{D}$, with metric $g^{MN}$, which is compactified to $ M^{4}
\times S/R$ with $S/R$ a coset space. The metric is assumed to
have the form
\begin{equation}
g^{MN}=
\left[\begin{array}{cc}\eta^{\mu\nu}&0\\0&-g^{ab}\end{array}
\right],
\end{equation}
where $\eta^{\mu\nu}= diag(1,-1,-1,-1)$ and $g^{ab}$ is the coset
space metric. The requirement that transformations of the fields
under the action of the symmetry group of $S/R$ are compensated by
gauge transformations lead to certain constraints on the fields.
The solution of these constraints provides us with the
four-dimensional unconstrained fields as well as with the gauge
invariance that remains in the theory after dimensional reduction.
Therefore a potential unification of all low energy interactions,
gauge, Yukawa and Higgs is achieved, which was the first
motivation of this framework.

It is interesting to note that the fields obtained using the CSDR
approach are the first terms in the expansion of the
$D$-dimensional fields in harmonics of the internal space $B$. The
effective field theories resulting from compactification of higher
dimensional theories contain also towers of massive higher
harmonics (Kaluza-Klein) excitations, whose contributions at the
quantum level alter the behaviour of the running couplings from
logarithmic to power \cite{Taylor}. As a result the traditional
picture of unification of couplings may change drastically
\cite{Dienes}. Higher dimensional theories have also been studied
at the quantum level using the continuous Wilson renormalization
group \cite{Kubo} which can be formulated in any number of
space-time dimensions with results in agreement with the treatment
involving massive Kaluza-Klein excitations.

Before we proceed with the description of the  CSDR scheme we need
to recall some facts about coset space geometry needed for
subsequent discussions. Complete reviews can be found in
\cite{Review,Castellani}.
\subsection{Coset Space Geometry.}
Assuming a $D$-dimensional spacetime $M^{D}$ with metric $g^{MN}$
given in eq.(1) it is instructive to explore further the geometry
of all coset spaces $S/R$.

We can divide the generators of $S$, $ Q_{A}$ in two sets : the
generators of $R$, $Q_{i}$ $(i=1, \ldots,dimR)$, and the
generators of $S/R$, $ Q_{a}$( $a=dimR+1 \ldots,dimS)$, and
$dimS/R=dimS-dimR =d$. Then the commutation relations for the
generators of $S$ are the following:
\begin{eqnarray}
\left[ Q_{i},Q_{j} \right] &=& f_{ij}^{\ \ k} Q_{k},\nonumber \\
\left[ Q_{i},Q_{a} \right]&=& f_{ia}^{\ \ b}Q_{b},\nonumber\\
\left[ Q_{a},Q_{b} \right]&=& f_{ab}^{\ \ i}Q_{i}+f_{ab}^{\ \
c}Q_{c} .
\end{eqnarray}
So $S/R$ is assumed to be a reductive but in general non-symmetric
coset space. When $S/R$ is symmetric, the $f_{ab}^{\ \ c}$ in
eq.(2) vanish. Let us call the coordinates of $M^{4} \times S/R$
space $z^{M}= (x^{m},y^{\alpha})$, where $\alpha$ is a curved
index of the coset,  $a$ is a tangent space index and $y$ defines
an element of $S$ which is a coset representative, $L(y)$. The
vielbein and the $R$-connection  are defined through the
Maurer-Cartan form which takes values in the Lie algebra of $S$ :
\begin{equation}
L^{-1}(y)dL(y) = e^{A}_{\alpha}Q_{A}dy^{\alpha} .
\end{equation}
Using eq.(3) we can compute that at the origin $y = 0$, $
e^{a}_{\alpha} = \delta^{a}_{\alpha}$ and $e^{i}_{\alpha} = 0$. A
connection on $S/R$ which is described by a connection-form
$\theta^{a}_{\ b}$, has in general torsion and curvature. In the
general case where torsion may be non-zero, we calculate first the
torsionless part $\omega^{a}_{\ b}$ by setting the torsion form
$T^{a}$ equal to zero,
\begin{equation}
T^{a} = de^{a} + \omega^{a}_{\ b} \wedge e^{b} = 0,
\end{equation}
while using the Maurer-Cartan equation,
\begin{equation}
de^{a} = \frac{1}{2}f^{a}_{\ bc}e^{b}\wedge e^{c} +f^{a}_{\
bi}e^{b}\wedge e^{i},
\end{equation}
we see that the condition of having vanishing torsion is solved by
\begin{equation}
\omega^{a}_{\ b}= -f^{a}_{\ ib}e^{i}-\frac{1}{2}f^{a}_{\ bc}e^{c}-
\frac{1}{2}K^{a}_{\ bc}e^{c},
\end{equation}
where $K^{a}_{\ bc}$ is symmetric in the indices $b,c$, therefore
$K^{a}_{\ bc}e^{c} \wedge e^{b}=0$. The $K^{a}_{\ bc}$ can be
found from the antisymmetry of $\omega^{a}_{\ b}$, $\omega^{a}_{\
b}g^{cb}=-\omega^{b}_{\ c}g^{ca}$, leading to
\begin{equation}
K_{\ \ bc}^{a}=g^{ad}(g_{be}f_{dc}^{\ \ e}+g_{ce}f_{db}^{\ \ e}).
\end{equation}
In turn $\omega^{a}_{\ b}$ becomes
\begin{equation}
\omega^{a}_{\ b}= -f^{a}_{\ ib}e^{i}-D_{\ \ bc}^{a}e^{c},
\end{equation}
where $$ D_{\ \ bc}^{a}=\frac{1}{2}g^{ad}[f_{db}^{\ \
e}g_{ec}+f_{cb}^{\ \ e} g_{de}- f_{cd}^{\ \ e}g_{be}].$$ The $D$'s
can be related to $f$'s by a rescaling \cite{Review}: $$ D^{a}_{\
bc}=(\lambda^{a}\lambda^{b}/\lambda^{c})f^{a}_{\ bc},$$ where the
$\lambda$'s depend on the coset radii. Note that in general the
rescalings change the antisymmetry properties of $f$'s, while
 in the case of equal radii $D^{a}_{\ bc}=\frac{1}{2}f^{a}_{\ bc}$.
Note also that the connection-form $\omega^{a}_{\ b}$ is
$S$-invariant. This means that parallel transport commutes with
the $S$ action \cite{Castellani}. Then the most general form of an
$S$-invariant connection on $S/R$ would be
\begin{equation}
\omega^{a}_{\ b} = f_{\ \ ib}^{a}e^{i}+J_{\ \ cb}^{a}e^{c},
\end{equation}
with $J$ an $R$-invariant tensor, i.e. $$ \delta J_{cb}^{\ \ a}
=-f_{ic}^{\ \ d}J_{db}^{\ \ a}+ f_{id}^{\ \ a}J_{cb}^{\ \
d}-f_{ib}^{\ \ d}J_{cd}^{\ \ a}=0. $$ This condition is satisfied
by the $D$'s as can be proven using the Jacobi identity.

 In the case of non-vanishing torsion we have
\begin{equation}
T^{a} = de^{a} + \theta^{a}_{\ b} \wedge e^{b},
\end{equation}
where $$\theta^{a}_{\ b}=\omega^{a}_{\ b}+\tau^{a}_{\ b},$$ with
\begin{equation}
\tau^{a}_{\ b} = - \frac{1}{2} \Sigma^{a}_{\ bc}e^{c},
\end{equation}
while the contorsion $ \Sigma^{a}_{\ \ bc} $ is given by
\begin{equation}
\Sigma^{a}_{\ \ bc} = T^{a}_{\ \ bc}+T_{bc}^{\ \ a}-T_{cb}^{\ \ a}
\end{equation}
in terms of the torsion components $ T^{a}_{\ \ bc} $. Therefore
in general the connection-form $ \theta^{a}_{\ b}$ is
\begin{equation}
\theta^{a}_{\ b} = -f^{a}_{\ ic}e^{i} -(D^{a}_{\
bc}+\frac{1}{2}\Sigma^{a}_{\ bc})e^{c}= -f^{a}_{\
ic}e^{i}-G^{a}_{\ bc}e^{c}.
\end{equation}
The natural choice of torsion which would generalize the case of
equal radii \cite{DLust,Gavrilik,Batakis}, $T^{a}_{\ bc}=\eta
f^{a}_{\ bc}$ would be $T^{a}_{\ bc}=2\tau D^{a}_{\ bc}$ except
that the $D$'s do not have the required symmetry properties.
Therefore we must define $\Sigma$ as a combination of $D$'s which
makes $\Sigma$ completely antisymmetric  and $S$-invariant
according to the definition given above. Thus we are led to the
definition
\begin{equation}
\Sigma_{abc} \equiv 2\tau(D_{abc}+D_{bca}-D_{cba}).
\end{equation}
 In this general case the Riemann curvature two-form is given by
\cite{Review}, \cite{Batakis}:
\begin{equation}
R^{a}_{\ b}=[-\frac{1}{2}f_{ib}^{\ a}f_{de}^{\ i}-
\frac{1}{2}G_{cb}^{\ a}f_{de}^{\ c}+ \frac{1}{2}(G_{dc}^{\
a}G_{eb}^{\ c}-G_{ec}^{\ a}G_{db}^{\ c})]e^{d} \wedge e^{e},
\end{equation}
whereas the Ricci tensor $R_{ab}=R^{d}_{\ adb}$ is
\begin{equation}
R_{ab}=G_{ba}^{\ c}G_{dc}^{\ d}-G_{bc }^{\ d}G_{da }^{\ c}-G_{ca
}^{\ d}f_{db }^{\ c}-f_{ia }^{\ d}f_{db }^{\ i}.
\end{equation}
By choosing the parameter $\tau$ to be equal to zero we can obtain
the { \it Riemannian connection} $\theta_{R \ \ b}^{\ a}$. We can
also define the { \it canonical connection} by adjusting the radii
and $\tau$ so that the connection form is $\theta_{C \ \ b}^{\ a}
= -f^{a}_{\ bi}e^{i}$, i.e. an $R$-gauge field \cite{DLust}. The
adjustments should be such that $G_{abc}=0$. In the case of
$G_{2}/SU(3)$ where the metric is $g_{ab}=a\delta_{ab}$, we have $
G_{abc}=\frac{1}{2}a(1+3\tau)f_{abc}$ and in turn $\tau=
-\frac{1}{3}$. In the case of $Sp(4)/(SU(2) \times
U(1))_{non-max.}$, where the metric is $g_{ab}=diag(a,a,b,b,a,a)$,
we have to set $a=b$ and then $\tau=- \frac{1}{3}$ to obtain the
canonical connection. Similarly in the case of $SU(3)/(U(1) \times
U(1))$, where the metric is $g_{ab}=diag(a,a,b,b,c,c)$, we should
set $a=b=c$ and take $\tau=- \frac{1}{3}$. By analogous
adjustments we can set the Ricci tensor equal to zero
\cite{DLust}, thus defining a {\it Ricci flattening connection}.
\subsection{Reduction of a $D$-dimensional Yang-Mills-Dirac
Lagrangian.}
 The group $S$ acts as a symmetry group on the extra
coordinates. The CSDR scheme demands that an $S$-transformation
of the extra $d$ coordinates is a gauge transformation of the
fields that are defined on $M^{4}\times S/R$,  thus a gauge
invariant Lagrangian written on this space is independent of the
extra coordinates.

To see this in detail we consider a $D$-dimensional
Yang-Mills-Dirac theory with gauge group $G$ defined on a
manifold $M^{D}$ which as stated will be compactified to
$M^{4}\times S/R$, $D=4+d$, $d=dimS-dimR$:
\begin{equation}
A=\int d^{4}xd^{d}y\sqrt{-g}\Bigl[-\frac{1}{4}
Tr\left(F_{MN}F_{K\Lambda}\right)g^{MK}g^{N\Lambda}
+\frac{i}{2}\overline{\psi}\Gamma^{M}D_{M}\psi\Bigr] ,
\end{equation}
where
\begin{equation}
D_{M}= \partial_{M}-\theta_{M}-A_{M},
\end{equation}
with
\begin{equation}
\theta_{M}=\frac{1}{2}\theta_{MN\Lambda}\Sigma^{N\Lambda}
\end{equation}
the spin connection of $M^{D}$, and
\begin{equation}
F_{MN}
=\partial_{M}A_{N}-\partial_{N}A_{M}-\left[A_{M},A_{N}\right],
\end{equation}
where $M$, $N$ run over the $D$-dimensional space. The fields
$A_{M}$ and $\psi$ are, as explained, symmetric in the sense that
any transformation under symmetries of $S/R$  is compensated by
gauge transformations. The fermion fields can be in any
representation $F$ of $G$ unless a further symmetry such as
supersymmetry is required. So let $\xi_{A}^{\alpha}$, $A
=1,\ldots,dimS$, be the Killing vectors which generate the
symmetries of $S/R$ and $W_{A}$ the compensating gauge
transformation associated with $\xi_{A}$. Define next the
infinitesimal coordinate transformation as $\delta_{A} \equiv
L_{\xi_{A}}$, the Lie derivative with respect to $\xi$, then we
have for the scalar,vector and spinor fields,
\begin{eqnarray}
\delta_{A}\phi&=&\xi_{A}^{\alpha}\partial_{\alpha}\phi=D(W_{A})\phi,
\nonumber \\
\delta_{A}A_{\alpha}&=&\xi_{A}^{\beta}\partial_{\beta}A_{\alpha}+\partial_{\alpha}
\xi_{A}^{\beta}A_{\beta}=\partial_{\alpha}W_{A}-[W_{A},A_{\alpha}],
\\
\delta_{A}\psi&=&\xi_{A}^{\alpha}\psi-\frac{1}{2}G_{Abc}\Sigma^{bc}\psi=
D(W_{A})\psi. \nonumber
\end{eqnarray}
$W_{A}$ depend only on internal coordinates $y$ and $D(W_{A})$
represents a gauge transformation in the appropriate
representation of the fields. $G_{Abc}$ represents a tangent space
rotation of the spinor fields. The variations $\delta_{A}$
satisfy, $[\delta_{A},\delta_{B}]=f_{AB}^{\\C}\delta_{C}$ and
lead to the following consistency relation for $W_{A}$'s,
\begin{equation}
\xi_{A}^{\alpha}\partial_{\alpha}W_{B}-\xi_{B}^{\alpha}\partial_{\alpha}
W_{A}-\left[W_{A},W_{B}\right]=f_{AB}^{\ \ C}W_{C}.
\end{equation}
 Furthermore the W's themselves transform under a gauge
transformation \cite{Review} as,
\begin{equation}
\widetilde{W}_{A} = gW_{A}g^{-1}+(\delta_{A}g)g^{-1}.
\end{equation}
Using eq.(23) and the fact that the Lagrangian is independent of
$y$ we can do all calculations at $y=0$ and choose a gauge where
$W_{a}=0$.

The detailed analysis of the constraints (21) given in
refs.\cite{Manton,Review} provides us with the four-dimensional
unconstrained fields as well as with the gauge invariance that
remains in the theory after dimensional reduction. Here we give
the results. The components $A_{\mu}(x,y)$ of the initial gauge
field $A_{M}(x,y)$ become, after dimensional reduction, the
four-dimensional gauge fields and furthermore they are independent
of $y$. In addition one can find that they have to commute with
the elements of the $R_{G}$ subgroup of $G$. Thus the
four-dimensional gauge group $H$ is the centralizer of $R$ in $G$,
$H=C_{G}(R_{G})$. Similarly, the $A_{\alpha}(x,y)$ components of
$A_{M}(x,y)$ denoted by $\phi_{\alpha}(x,y)$ from now on, become
scalars at four dimensions. These fields transform under $R$ as a
vector $v$, i.e.
\begin{eqnarray}
S &\supset& R \nonumber \\
adjS &=& adjR+v.
\end{eqnarray}
Moreover $\phi_{\alpha}(x,y)$ act as an intertwining operator
connecting induced representations of $R$ acting on $G$ and $S/R$.
This implies, exploiting Schur's lemma, that the transformation
properties of the fields $\phi_{\alpha}(x,y)$ under $H$ can be
found if we express the adjoint representation of $G$ in terms of
$R_{G} \times H$ :
\begin{eqnarray}
G &\supset& R_{G} \times H \nonumber \\
 adjG &=&(adjR,1)+(1,adjH)+\sum(r_{i},h_{i}).
\end{eqnarray}
Then if $v=\sum s_{i}$, where each $s_{i}$ is an irreducible
representation of $R$, there survives an $h_{i}$ multiplet for
every pair $(r_{i},s_{i})$, where $r_{i}$ and $s_{i}$ are
identical irreducible representations of $R$.

Turning next to the fermion fields
\cite{Review,Slansky,Chapline,Palla} similarly to scalars, they
act as intertwining operators between induced representations
acting on $G$ and the tangent space of $S/R$, $SO(d)$. Proceeding
along similar lines as in the case of scalars to obtain the
representation of $H$ under which the four-dimensional fermions
transform, we have to decompose the representation $F$ of the
initial gauge group in which the fermions are assigned under
$R_{G} \times H$, i.e.
\begin{equation}
F= \sum (t_{i},h_{i}),
\end{equation}
and the spinor of $SO(d)$ under $R$
\begin{equation}
\sigma_{d} = \sum \sigma_{j}.
\end{equation}
Then for each pair $t_{i}$ and $\sigma_{i}$, where $t_{i}$ and
$\sigma_{i}$ are identical irreducible representations there is an
$h_{i}$ multiplet of spinor fields in the four-dimensional theory.
In order however  to obtain chiral fermions in the effective
theory we have to impose further requirements. We first impose the
Weyl condition in $D$ dimensions. In $D = 4n+2$ dimensions which
is the case at hand, the decomposition of the left handed, say
spinor under $SU(2) \times SU(2) \times SO(d)$ is
\begin{equation}
\sigma _{D} = (2,1,\sigma_{d}) + (1,2,\overline{\sigma}_{d}).
\end{equation}
So we have in this case the decompositions
\begin{equation}
\sigma_{d} = \sum \sigma_{k},~\overline{\sigma}_{d}= \sum
\overline{\sigma}_{k}.
\end{equation}
Let us start from a vector-like representation $F$ for the
fermions. In this case each term $(t_{i},h_{i})$ in eq.(26) will
be either self-conjugate or it will have a partner $(
\overline{t}_{i},\overline{h}_{i} )$. According to the rule
described in eqs.(26), (27) and considering $\sigma_{d}$ we will
have in four dimensions left-handed fermions transforming as $
f_{L} = \sum h^{L}_{k}$. It is important to notice that since
$\sigma_{d}$ is non self-conjugate, $f_{L}$ is non self-conjugate
too. Similarly from $\overline{\sigma}_{d}$ we will obtain the
right handed representation $ f_{R}= \sum \overline{h}^{R}_{k}$
but as we have assumed that $F$ is vector-like,
$\overline{h}^{R}_{k}\sim h^{L}_{k}$. Therefore there will appear
two sets of Weyl fermions with the same quantum numbers under $H$.
This is already a chiral theory, but still one can go further and
try to impose the Majorana condition in order to eliminate the
doubling of the fermionic spectrum. We should remark now that if
we had started with $F$ complex, we should have again a chiral
theory since in this case $\overline{h}^{R}_{k}$ is different from
$h^{L}_{k}$  $(\sigma_{d}$ non self-conjugate). Nevertheless
starting with $F$ vector-like is much more appealing and will be
used in the following along with the Majorana condition. The
Majorana condition can be imposed in $D = 2,3,4+8n$ dimensions and
is given by $\psi = C\overline\psi^{T}$, where $C$ is the
$D$-dimensional charge conjugation matrix. Majorana and Weyl
conditions are compatible in $D=4n+2$ dimensions. Then in our case
if we start with Weyl-Majorana spinors in $D=4n+2$ dimensions we
force $f_{R}$ to be the charge conjugate to $f_{L}$, thus arriving
in a theory with fermions only in $f_{L}$. Furthermore if $F$ is
to be real, then we have to have $D=2+8n$, while for $F$
pseudoreal $D=6+8n$.

Starting with an anomaly free theory in higher dimensions, in
ref.\cite{Witten} was given the condition that has to be
fulfilled in order to obtain anomaly free theories in four
dimensions after dimensional reduction. The condition restricts
the allowed embeddings of $R$ into $G$ \cite{Pilch,Review}. For
$G=E_{8}$ in ten dimensions the condition takes the form
\begin{equation}
l(G) = 60,
\end{equation}
where $l(G)$ is the sum over all indices of the $R$
representations appearing in the decomposition of the $248$
representation of $E_{8}$ under $ E_{8} \supset R \times H$. The
normalization is such that the vector representation in eq.(24)
which defines the embedding of $R$ into $SO(6)$, has index two.
\subsection{The Four-Dimensional Theory.}
Next let us obtain the four-dimensional effective action. Assuming
that the metric is block diagonal, taking into account all the
constraints and integrating out the extra coordinates we obtain in
four dimensions the following Lagrangian :
\begin{equation}
A=C \int d^{4}x \biggl( -\frac{1}{4} F^{t}_{\mu
\nu}{F^{t}}^{\mu\nu}+\frac{1}{2}(D_{\mu}\phi_{\alpha})^{t}
(D^{\mu}\phi^{\alpha})^{t}
+V(\phi)+\frac{i}{2}\overline{\psi}\Gamma^{\mu}D_{\mu}\psi-\frac{i}{2}
\overline{\psi}\Gamma^{a}D_{a}\psi\biggr),
\end{equation}
where $D_{\mu} = \partial_{\mu} - A_{\mu}$ and $D_{a}=
\partial_{a}- \theta_{a}-\phi_{a}$ with  $\theta_{a}=
\frac{1}{2}\theta_{abc}\Sigma^{bc}$ the connection of the coset
space, while $C$ is the volume of the coset space. The potential
$V(\phi)$ is given by:
\begin{equation}
V(\phi) = - \frac{1}{4} g^{ac}g^{bd}Tr( f _{ab}^{C}\phi_{C} -
[\phi_{a},\phi_{b}] ) (f_{cd}^{D}\phi_{D} - [\phi_{c},\phi_{d}] )
,
\end{equation}
where, $A=1,\ldots,dimS$ and $f$ ' s are the structure constants
appearing in the commutators of the generators of the Lie algebra
of S. The expression (32) for $V(\phi)$ is only formal because
$\phi_{a}$ must satisfy the constraints coming from eq.(21),
\begin{equation}
f_{ai}^{D}\phi_{D} - [\phi_{a},\phi_{i}] = 0,
\end{equation}
where the $\phi_{i}$ generate $R_{G}$. These constraints imply
that some components $\phi_{a}$'s are zero, some are constants and
the rest can be identified with the genuine Higgs fields. When
$V(\phi)$ is expressed in terms of the unconstrained independent
Higgs fields, it remains a quartic polynomial which is invariant
under gauge transformations of the final gauge group $H$, and its
minimum determines the vacuum expectation values of the Higgs
fields \cite{Vinet,Harnad,Farakos}. The minimization of the
potential is in general a difficult problem. If however $S$ has an
isomorphic image $S_{G}$ in $G$ which contains $R_{G}$ in a
consistent way then it is possible to allow the $\phi_{a}$ to
become generators of $S_{G}$. That is $\overline{\phi}_{a} =
<\phi^{i}>Q_{ai} = Q_{a}$ with $<\phi^{i}>Q_{ai}$ suitable
combinations of $G$ generators, $Q_{a}$ a generator of $S_{G}$
and $a$ is also a coset-space index. Then
\begin{eqnarray*}
\overline{F}_{ab}&=&f_{ab}^{\ \ i}Q_{i}+f_{ab}^{\ \
c}\overline{\phi}_{c}-[\overline{\phi}_{a},\overline{\phi}_{b}]\\
&=& f_{ab}^{\ \ i}Q_{i}+ f_{ab}^{\ \ c}Q_{c}- [Q_{a},Q_{b}] = 0
\end{eqnarray*}
because of the commutation relations of $S$. Thus we  have proven
that $V(\phi=\overline{\phi})=0$ which furthermore is the minimum,
because $V$ is positive definite. Furthermore, the
four-dimensional gauge group $H$ breaks further by these non-zero
vacuum expectation values of the Higgs fields to the centralizer
$K$ of the image of $S$ in $G$, i.e. $K=C_{G}(S)$
\cite{Review,Vinet,Harnad,Farakos}. This can been seen if we
examine a gauge transformation of $\phi_{a}$ by an element $h$ of
$H$. Then we have $$ \phi_{a} \rightarrow h\phi_{a}h^{-1}, h \in
H $$ We note that the v.e.v. of the Higgs fields is gauge
invariant for the set of $h$'s that commute with $S$. That is $h$
belongs to a subgroup $K$ of $H$ which is the centralizer of
$S_{G}$ in $G$.

More generally it can be proven \cite{Review} that dimensional
reduction over a symmetric coset space always gives a potential of
spontaneous breaking form. Note that in this case the potential
acquires the form,
\begin{equation}
V(\phi)=-\frac{1}{4}g^{ac}g^{bd}Tr(f_{ab}^{\ \
i}J_{i}-[\phi_{a},\phi_{b}])(f_{cd}^{\ \
j}J_{j}-[\phi_{a},\phi_{b}]).
\end{equation}
since the structure constants $f_{ab}^{\ \ c}$ are equal to zero.
Next we decompose the adjoint representation of $S$ under $R$,
\begin{eqnarray}
S &\supset& R \nonumber \\
adjS &=& adjR+\Sigma(s_{a}+\overline{s}_{a}),
\end{eqnarray}
and introduce the generators of the coset,
\begin{equation}
Q_{S}=(Q_{i},Q_{s_{a}},Q_{\overline{s}_{a}}),
\end{equation}
where $Q_{i}$ correspond to $R$ and $Q_{s_{a}}$ and
$Q_{\overline{s}_{a}}$ to $s_{a}$ and $\overline{s}_{a}$. With
this notation and using the complex metric $g^{i\overline{j}}$ the
potential (34) can be rewritten as
\begin{eqnarray}
V(\phi)=-\frac{1}{2}g^{s_{a}\overline{s}_{a}}g^{t_{a}\overline{t}_{a}}
Tr(f_{s_{a}t_{a}}^{\ \
i}J_{i}-[\phi_{s_{a}},\phi_{t_{a}}])(f_{\overline{s}_{a}\overline{t}_{a}}^{\
\ j}J_{j}-[\phi_{\overline{s}_{a}},\phi_{\overline{t}_{a}}])
\nonumber
\\
-\frac{1}{2}g^{s_{a}\overline{s}_{a}}g^{t_{a}\overline{t}_{a}}Tr(f_{s_{a}\overline{t}_{a}}^{\
\
i}J_{i}-[\phi_{s_{a}},\phi_{\overline{t}_{a}}])(f_{\overline{s}_{a}t_{a}}^{\
\ j}J_{j}-[\phi_{\overline{s}_{a}},\phi_{t_{a}}]).
\end{eqnarray}
Note that the structure constants involved in the first and the
second parentheses inside the traces in eq.(37) are of opposite
sign, since they appear in the commutator of conjugate generators.
The same is true for the commutator of two $\phi$ fields, since
they are actually expressed as linear combinations of the gauge
group generators; if $\phi_{s_{a}}$ is connected to one generator
then $\phi_{\overline{s}_{a}}$ will be connected to its conjugate
generator. As a result, terms involving two $J_{i}$ will be
constant positive terms, terms with one $J_{i}$ and a $\phi$
commutator  will be negative mass terms, and finally terms
involving two $\phi$  commutators will be quatric positive terms.
This result remains unaltered if the more general case is
considered, where the vector of the coset $S/R$ decomposed under
$R$ contains also real representations. So in conclusion the
potential obtained from the dimensional reduction of a gauge
theory over symmetric coset spaces is always of a spontaneous
breaking form.

In the fermion part of the Lagrangian the first term is just the
kinetic term of fermions, while the second is the Yukawa term
\cite{Kapetanakis}. Note that since $\psi$ is a Majorana-Weyl
spinor in ten dimensions the representation in which the fermions
are assigned under the gauge group must be real. The last term in
eq.(31) can be written as
\begin{equation}
L_{Y}= -\frac{i}{2}\overline{\psi}\Gamma^{a}(\partial_{a}-
\frac{1}{2}f_{ibc}e^{i}_{\gamma}e^{\gamma}_{a}\Sigma^{bc}-
\frac{1}{2}G_{abc}\Sigma^{bc}- \phi_{a}) \psi \nonumber \\
=\frac{i}{2}\overline{\psi}\Gamma^{a}\nabla_{a}\psi+
\overline{\psi}V\psi ,
\end{equation}
where
\begin{eqnarray}
\nabla_{a}& =& - \partial_{a} +
\frac{1}{2}f_{ibc}e^{i}_{\gamma}e^{\gamma}_{a}\Sigma^{bc} + \phi_{a},\\
 V&=&\frac{i}{4}\Gamma^{a}G_{abc}\Sigma^{bc},
\end{eqnarray}
and we have used the full connection with torsion \cite{Batakis}
given by
\begin{equation}
\theta_{\ \ c b}^{a} = - f_{\ \
ib}^{a}e^{i}_{\alpha}e^{\alpha}_{c}-(D_{\ \ cb}^{a} +
\frac{1}{2}\Sigma_{\ \ cb}^{a}) = - f_{\ \
ib}^{a}e^{i}_{\alpha}e^{\alpha}_{c}- G_{\ \ cb}^{a}
\end{equation}
with
\begin{equation}
D_{\ \ cb}^{a} = g^{ad}\frac{1}{2}[f_{db}^{\ \ e}g_{ec} + f_{
cb}^{\ \ e}g_{de} - f_{cd}^{\ \ e}g_{be}]
\end{equation}
and
\begin{equation}
\Sigma_{abc}= 2\tau(D_{abc} +D_{bca} - D_{cba}).
\end{equation}
 We have already noticed that the CSDR constraints tell us that
$\partial_{a}\psi= 0$. Furthermore we can consider the Lagrangian
at the point $y=0$, due to its invariance under
$S$-transformations, and as we mentioned $e^{i}_{\gamma}=0$ at
that point. Therefore eq.(39) becomes just $\nabla_{a}= \phi_{a}$
and the term $\frac{i}{2}\overline{\psi}\Gamma^{a}\nabla_{a}\psi $
in eq.(38) is exactly the Yukawa term.

Let us examine now the last term appearing in eq.(38). One can
show easily that the operator $V$ anticommutes with the
six-dimensional helicity operator \cite{Review}. Furthermore one
can show that $V$ commutes with the $T_{i}=
-\frac{1}{2}f_{ibc}\Sigma^{bc}$ ($T_{i}$ close the $R$-subalgebra
of $SO(6)$). In turn we can draw the conclusion, exploiting
Schur's lemma, that the non-vanishing elements of $V$ are only
those which appear in the decomposition of both $SO(6)$ irreps $4$
and $\overline{4}$, e.g. the singlets. Since this term is of pure
geometric nature, we reach the conclusion that the singlets in $4$
and $\overline{4}$ will acquire large geometrical masses, a fact
that has serious phenomenological implications. In supersymmetric
theories defined in higher dimensions, it means that the gauginos
obtained in four dimensions after dimensional reduction receive
masses comparable to the compactification scale. However as we
shall see in the next sections this result changes in presence of
torsion. We note that for symmetric coset spaces the $V$ operator
is absent because $f_{ab}^{c}$ are vanishing by definition in that
case.
\section{Supersymmetry Breaking by Dimensional Reduction over Coset Spaces.}
 Recently a lot of interest has been triggered by the
possibility that superstrings can be defined at the TeV scale
\cite{Anton}. The string tension became an arbitrary parameter and
can be anywhere below the Planck scale and as low as TeV. The main
advantage of having the string tension at the TeV scale, besides
the obvious experimental interest, is that it offers an automatic
protection to the gauge hierarchy \cite{Anton}, alternative to low
energy supersymmetry \cite{Dim}, or dynamical electroweak symmetry
breaking \cite{Fahri,Marciano,Trianta}. However the only vacua of
string theory free of all pathologies are supersymmetric. Then
the original supersymmetry of the theory, not being necessary in
four dimensions, could be broken by the dimensional reduction
procedure.

The weakly coupled ten-dimensional $E_{8} \times E_{8}$
supersymmetric gauge theory is one of the few to posses  the
advantage of anomaly freedom \cite{Green} and has been extensively
used in efforts to describe quantum gravity along with the
observed low energy interactions in the heterotic string framework
\cite{Theisen}. In addition its strong coupling limit  provides an
interesting example of the realization of the brane picture, i.e.
$E_{8}$ gauge fields and matter live on the two ten-dimensional
boundaries, while gravitons propagate in the eleven-dimensional
bulk \cite{Horava}.

In the following we shall be reducing a supersymmetric
ten-dimensional gauge theory based on $E_{8}$ over coset spaces
and examine the consequences of the resulting four-dimensional
theory mostly as far as supersymmetry breaking is concerned.
\subsection{Supersymmetry Breaking by Dimensional Reduction over
Symmetric Coset spaces.} Let us first examine the reduction of the
ten dimensional ${\cal N} = 1$ supersymmetric $E_{8}$ gauge theory
over the symmetric coset spaces of table 1.\\

\vspace{.1in}

{\it { \bf a. Reduction of $G=E_{8}$ over $B=SO(7)/SO(6)$}}\\

First we review the reduction of $E_{8}$ over the 6-sphere
\cite{Review,Chapline,Pman}. In that case $B=SO(7)/SO(6)$, $D=10$
and the Weyl-Majorana fermions belong in the adjoint of G. The
embedding of $R=SO(6)$ in $E_{8}$ is suggested by the
decomposition $$E_{8} \supset SO(6) \times SO(10) $$
\begin{equation}
248 = (15,1)+(1,45)+(6,10)+(4,16)+(\overline{4},\overline{16}).
\end{equation}
The $R=SO(6)$ content of the vector and spinor of $SO(7 )/SO(6)$
are $6$ and $4$ respectively. The condition that guarantees the
anomaly freedom of the four-dimensional theory given in eq.(30) is
satisfied and the resulting gauge group is
$H=C_{E_{8}}(SO(6))=SO(10)$. According to the CSDR rules (24),(25)
and (26),(27) the surviving scalars in four dimensions transform
as a 10-plet under the gauge group $SO(10)$, while the surviving
fermions provide the four-dimensional theory with a $16_{L}$ and a
$\overline{16}_{R}$ which are identified by the Weyl-Majorana
condition. Concerning supersymmetry obviously any sign of the
supersymmetry of the original theory has disappeared in the
process of dimensional reduction. On the other hand the
four-dimensional theory is a GUT with fermions in a multiplet
which is appropriate to describe quarks and leptons (including a
right-handed neutrino). Finally since the $SO(7)$ has an
isomorphic image in $E_{8}$, according to the theorem discussed in
the subsection 2.3, the $SO(10)$ breaks further due to the v.e.v.
of the $10$-plet Higgs down to $C_{E_{8}}(SO(7))=SO(9)$. Therefore
the scalar field content of the four-dimensional theory is
appropriate for the electroweak symmetry breaking but not for the
GUT breaking.\\

\vspace{.1in}

{\it{\bf b. Reduction of $G=E_{8}$ over $B=SU(4)/SU(3) \times
U(1).$}}\\

In this case $G=E_{8}$ and $S/R=SU(4)/SU(3) \times U(1)$. The
embedding of $R=SU(3) \times U(1)$ is determined by the following
decomposition $$E_{8} \supset SU(3) \times U(1) \times SO(10)$$
\begin{equation}
248=
(1_{0}+3_{-4}+\overline{3}_{4}+8_{0},1)+(1_{0},45)+(3_{2}+\overline{3}_{-2},10)
+(1_{3}+3_{-1},16)+(1_{-3}+3_{1},\overline{16}),
\end{equation}
where the $SU(3) \times U(1)$ is the maximal subgroup of $SO(6)
\approx SU(4)$ appearing in the decomposition (44). The $R$ is
chosen to be identified with the $SU(3) \times U(1)$ of the above
decomposition. Therefore the resulting four-dimensional gauge
group is $H=C_{E_{8}}(SU(3) \times U(1))=SO(10) \times U(1)$ (The
$U(1)$ appears since the $U(1)$ in $R$ centralizes itself). The
$R=SU(3) \times U(1)$ content of $SU(4)/SU(3) \times U(1)$ vector
and spinor can be read from table 1 and are
$3_{-2}+\overline{3}_{2}$ and $1_{3}+3_{-1}$ respectively.
Therefore we find that the surviving scalars in four dimensions
transform as $10_{2}$ and $10_{-2}$, while the four-dimensional
fermions transform as $16_{3}$ and $16_{-1}$ under the
four-dimensional gauge group $H=SO(10) \times U(1)$. Again there
is no sign in four dimensions of the original supersymmetry.\\

\vspace{.1in}

{\it { \bf c. Reduction of $G=E_{8}$ over $B=Sp(4)/(SU(2) \times
U(1))_{max.}$}}
\\

Next we choose $G=E_{8}$ and $S/R=Sp(4)/(SU(2) \times U(1))_{max.}
$. The embedding of $R=(SU(2) \times U(1))_{max}$ is determined by
the decomposition (45) when the $SU(2)$ is chosen to be the
maximal subgroup of $SU(3)$. In that case the decomposition of the
$248$ of $E_{8}$ is the following
$$E_{8} \supset SU(2) \times U(1) \times SO(10)$$
\begin{eqnarray}
248 =
(1,1)_{0}+(3,1)_{-4}+(3,1)_{4} \nonumber \\
+(3,1)_{0}+(5,1)_{0}+(1,45)_{0} \nonumber \\ + (3,10)_{2} +
(3,10)_{-2} + (1,16)_{3} \nonumber \\ +
(3,16)_{-1}+(1,\overline{16})_{-3}+(3,\overline{16})_{1}.
\end{eqnarray}
From the decomposition (46) is clear that the four-dimensional
gauge group is $H=C_{E_{8}}((SU(2) \times U(1))_{max.})=SO(10)
\times U(1)$. The $R=(SU(2) \times U(1))_{max.}$ content of
$Sp(4)/(SU(2) \times U(1))_{max.}$ vector and spinor according to
table 1 are $3_{-2}+3_{2}$ and $1_{3}+3_{-1}$ respectively under
$R$. Therefore the particle content of the four-dimensional theory
is a set of $10_{2}$, $10_{-2}$ scalars and a set of $16_{3}$,
$16_{-1}$ left handed spinors. Once more no sign of the original
supersymmetry is left in the spectrum of the four-dimensional
theory.\\

\vspace{.1in}

{\it{\bf d. Reduction of $G=E_{8}$ over $B=Sp(4) \times
SU(2)/SU(2) \times SU(2) \times U(1)$.}}\\

Next we choose again $G=E_{8}$ while $S/R=Sp(4) \times SU(2)/SU(2)
\times SU(2) \times U(1)$. The embedding of $R=SU(2) \times SU(2)
\times U(1)$ is given by the decomposition $$E_{8} \supset SU(2)
\times SU(2) \times U(1) \times SO(10)$$
\begin{eqnarray}
248=(3,1,1)_{0}+(1,3,1)_{0}+(1,2,16)_{-1}+(1,2,\overline{16})_{1}
\nonumber \\ +(1,1,1)_{0}+(1,1,45)_{0}+(1,1,10)_{2}+(1,1,10)_{-2}
\nonumber \\+(2,2,1)_{2}+(2,2,1)_{-2}+(2,2,10)_{0} \nonumber \\
+(2,1,16)_{1}+(2,1,\overline{16})_{-1},
\end{eqnarray}
where the $R=SU(2) \times SU(2) \times U(1)$ is the maximal
subgroup of $ SO(6) \approx SU(4)$ appearing in the decomposition
(44). The four-dimensional gauge group that survives after
dimensional reduction is $H=C_{E_{8}}(SU(2) \times SU(2) \times
U(1))=SO(10) \times U(1)$. According to table 1 the $R=SU(2)
\times SU(2) \times U(1)$ content of $Sp(4) \times SU(2)/SU(2)
\times SU(2) \times U(1)$ vector and spinor are
$(2,2)_{0}+(1,1)_{2}+(1,1)_{-2}$ and $(2,1)_{1}+(1,2)_{-1}$,
respectively. Therefore the scalar fields that survive in four
dimensions belong to $10_{0}$, $10_{2}$, $10_{-2}$ of $H=SO(10)
\times U(1)$. Similarly the surviving fermions in four dimensions
transform as $16_{1}$, $16_{-1}$ left-handed multiplets.\\

\vspace{.1in}

{\it{\bf e. Reduction of $G=E_{8}$ over $B=SU(3) \times
SU(2)/SU(2) \times U(1) \times U(1)$}}\\

Choosing $G=E_{8}$ and $S/R=SU(3) \times SU(2)/SU(2) \times U(1)
\times U(1)$ we have another interesting example. The embedding of
$R=SU(2) \times U(1) \times U(1)$ in $E_{8}$ is given by the
decomposition $$E_{8} \supset SU(2) \times U(1) \times U(1) \times
SO(10)$$
\begin{eqnarray}
248=(1,45)_{(0,0)}+(3,1)_{(0,0)}+(1,1)_{(0,0)}+(1,1)_{(0,0)}+(1,1)_{(2,0)}
\nonumber \\
+(1,1)_{(-2,0)}+(2,1)_{(1,2)}+(2,1)_{(-1,2)}+(2,1)_{(-1,-2)}+(2,1)_{(1,-2)}
\nonumber \\
+(1,10)_{(0,2)}+(1,10)_{(0,-2)}+(2,10)_{(1,0)}+(2,10)_{(-1,0)}
\nonumber \\ +(2,16)_{(0,1)}+(1,16)_{(1,-1)}+(1,16)_{(-1,-1)}
\nonumber \\
+(2,\overline{16})_{(0,-1)}+(1,\overline{16})_{(-1,1)}+(1,\overline{16})_
{(1,1)},
\end{eqnarray}
where the $R=SU(2) \times U(1) \times U(1)$ is identified with the
one appearing in the following decomposition of maximal subgroups
$$ SO(6) \supset SU(2) \times SU(2) \times U(1) \supset SU(2)
\times U(1) \times U(1)$$ and the $SU(2) \times SU(2) \times
U(1)$ in $E_{8}$ is the maximal subgroup of $SO(6)$ appearing in
the decomposition (44). We find that the four-dimensional gauge
group is $H=C_{E_{8}}(SU(2) \times U(1) \times U(1))=SO(10) \times
U(1) \times U(1)$. The vector and spinor content under $R$ of the
specific coset can be found in table 1. Choosing $a=b=1$ we find
that the scalar fields of the four-dimensional theory transform as
$10_{(0,2)}$, $10_{(0,-2)}$, $10_{(1,0)}$, $10_{(-1,0)}$ under
$H$. Also, we find that the fermions of the four-dimensional
theory are the following left-handed multiplets of $H$:
$16_{(-1,-1)}$, $16_{(1,-1)}$, $16_{(0,1)}$.\\

\vspace{.1in}

{\it { \bf f. Reduction  of $G=E_{8}$ over
$B=(SU(2)/U(1))^{3}$.}}\\

Last we examine the case with $G=E_{8}$ and
$S/R=(SU(2)/U(1))^{3}$. The $R=U(1)^{3}$ is chosen to be
identified with the three $U(1)$ subgroups of $SO(6)$ appearing in
the decomposition $$SO(6) \supset SU(2) \times SU(2) \times U(1)
\supset U(1) \times U(1) \times U(1),$$ where the $SO(6)$ is again
the one of the decomposition (44). Then we find the following
decomposition of $248$ of $E_{8}$,  $$E_{8} \supset U(1) \times
U(1) \times U(1) \times SO(10)$$
\begin{eqnarray}
248 =
1_{(0,0,0)}+1_{(0,0,0)}+1_{(0,0,0)}+1_{(0,0,\pm2)}+1_{(\pm4,\mp2,0)}\nonumber\\
+1_{(0,\pm3,\pm1)}+1_{(0,\mp3,\pm1)}+1_{(\pm4,\pm1,\pm1)}+1_{(\pm4,\pm1,\mp1)}\nonumber\\
+45_{(0,0,0)}+16_{(-3,0,0)}+\overline{16}_{(3,0,0)}+16_{(1,-2,0)}\nonumber\\
+\overline{16}_{(-1,2,0)}+10_{(-2,-2,0)}+\overline{10}_{(2,2,0)}\nonumber\\
+16_{(1,1,1)}+\overline{16}_{(-1,-1,-1)}+16_{(1,1,-1)}
+\overline{16}_{(-1,-1,1)}\nonumber\\+10_{(-2,1,1)}+\overline{10}_{(2,-1,-1)}
+10_{(-2,1,-1)}+\overline{10}_{(2,-1,1)}.
\end{eqnarray}
Therefore the four-dimensional gauge group is
$H=C_{E_{8}}(U(1)^{3})=SO(10)\times U(1)^{3} $. The $R=U(1)^{3}$
content of $(SU(2)/U(1))^{3}$ vector and spinor can be found in
table 1. With $a=b=c=1$ the vector and spinor become $(0,0,\pm
2)+(\pm 2,0,0)+(0,\pm 2,0)$ and $(1,1,1)+(-1,-1, 1)+(-1, 1,-1)+(1,
-1,-1)$ respectively and therefore the four-dimensional scalar
fields transform as singlets, while fermions transform as
left-handed $16_{(1,1,1)}$, and $16_{(1,1,-1)}$ under $H$.
\\

Note that in all above cases a - f the chosen embeddings of $R$ in
$G$ satisfy eq.(30) and therefore the the resulting
four-dimensional theories are anomaly free.

\vspace{.1in}

\subsection{Supersymmetry Breaking by Dimensional Reduction over
non-symmetric Coset Spaces } Next we start with the same theory in
ten dimensions but we reduce it over the three non-symmetric coset
spaces listed in table 2.\\

\vspace{.1in}

{\it{\bf a. Soft Supersymmetry Breaking by dimensional reduction
over $G_{2}/SU(3)$}}\\

First we choose $B= G_{2}/SU(3)$ \cite{Pman}. We use the
decomposition
\begin{eqnarray}
E_{8} &\supset& SU(3) \times E_{6}\nonumber \\ 248 &=& (8,1) +
(1,78) + (3,27) + (\overline{3},\overline{27})
\end{eqnarray}
and we choose $SU(3)$ to be identified with $R$. The $R=SU(3)$
content  of $G_{2}/SU(3)$ vector and spinor is $ 3 + \overline{3}$
and $1+3$ as can be read from table 2. The condition (30) for the
cancellation of anomalies is satisfied and the resulting
four-dimensional gauge group is $ H = C_{E_{8}}(SU(3)) = E_{6}$,
which contains fermion and complex scalar fields transforming as
78, 27 and 27 respectively. Therefore we obtain in four
dimensions a ${\cal N}=1$ supersymmetric anomaly free $E_{6}$
gauge theory with a vector superfield grouping gauge bosons and
fermions transforming according to the adjoint and a matter chiral
superfield grouping scalars and fermions in the fundamental of the
gauge group $E_{6}$. In addition a very interesting feature worth
stressing is that the ${\cal N}=1$ supersymmetry of the
four-dimensional theory is broken by soft terms. More precisely
the scalar soft terms appear in the potential of the theory and
the gaugino masses come from a geometric (torsion) term as already
stated.

We proceed by calculating these terms. In order to determine the
potential we begin by examining the decomposition of the adjoint
of the specific $S$ under $R$, i.e.
\begin{eqnarray}
G_{2} &\supset& SU(3) \nonumber \\ 14 &=& 8+3+\overline{3}.
\end{eqnarray}
Corresponding to this decomposition we introduce the generators of
$G_{2}$
\begin{equation}
Q_{G_{2}} = \{ Q^{a},Q^{\rho},Q_{\rho}\},
\end{equation}
where $a=1,\ldots,8$ correspond  to the $8$ of $SU(3)$, while
$\rho = 1,2,3$ correspond to $3$ or $\overline{3}$. Then according
to the decomposition (52), the non trivial commutation relations
of the generators of $G_{2}$ are given in table 3 of appendix A.

The potential of any theory reduced over $G_{2}/SU(3)$ can be
written in terms of the fields
\begin{equation}
\{ \phi^{a} , \phi^{\rho} ,\phi_{\rho} \},
\end{equation}
which correspond to the decomposition (52) of $G_{2}$ under
$SU(3)$. The $\phi_{a}$ are equal to the generators of the  $R$
subgroup. With the help of the commutation relations of Table 3 we
find that the potential of any theory reduced over $G_{2}/SU(3)$
is given by \cite{Review} :
\begin{eqnarray}
V(\phi)=\frac{8}{R_{1}^{4}}+\frac{4}{3R_{1}^{4}}Tr(\phi^{\rho}\phi_{\rho})
-\frac{1}{2R_{1}^{4}}(\lambda^{i})^{\rho}_{\sigma}Tr(J_{i}[\phi_{\rho},\phi^{\sigma}])
+\frac{1}{R_{1}^{4}}\sqrt{\frac{2}{3}}\epsilon^{\rho\sigma\tau}Tr(\phi_{\tau}[\phi_{\rho},\phi_{\sigma}])
\nonumber \\
 -\frac{1}{4R_{1}^{4}} Tr([\phi_{\rho},\phi_{\sigma}][\phi^{\rho},\phi^{\sigma}] +
[\phi^{\rho},\phi_{\sigma}][\phi_{\rho},\phi^{\sigma}]),
\end{eqnarray}
where the $R_{1}$ appearing in the denominator of various terms is
the radius of $G_{2}/SU(3)$. Then to proceed with our specific
choice $G=E_{8}$ we use the embedding (50) of $R=SU(3)$ in $E_{8}$
and divide accordingly the generators of $E_{8}$
\begin{equation}
Q_{E_{8}} = \{ Q^{a} , Q^{\alpha},Q^{i\rho},Q_{i\rho} \}
\end{equation}
with $a = 1,\ldots,8$,  $\alpha = 1,\ldots,78$, $i=1,\ldots,27$,
$\rho=1,2,3$. The non-trivial commutation relations of the
generators of $E_{8}$ according to the decomposition (55) are
given in table 4 of appendix A. Next we would like to solve the
constraints (33) which in the present case take the form $[
\phi^{a},\phi^{\rho}]=-(\lambda^{a})^{\rho}_{\sigma}\phi^{\sigma}$,
and examine the resulting four-dimensional potential in terms of
the unconstrained scalar fields $\beta$. The solutions of the
constraints in terms of the genuine Higgs fields are
\begin{equation}
\phi^{a}=Q^{a},\  \phi_{\rho}=R_{1}\beta^{i}Q_{i\rho},\
\phi^{\rho}=R_{1}\beta_{i}Q^{i\rho}.
\end{equation}
In turn we can express the Higgs potential in terms of the
genuine Higgs field $\beta$ and we find
\begin{equation}
V(\beta)= \frac{8}{R_{1}^{4}}- \frac{40}{3R_{1}^{2}}\beta^{2} -
\left[\frac{4}{R_{1}}d_{ijk}\beta^{i}\beta^{j}\beta^{k} + h.c
\right] +\beta^{i}\beta^{j}d_{ijk}d^{klm}\beta_{l}\beta_{m}+
\frac{11}{4}\sum_{\alpha}\beta^{i}(G^{\alpha})_{i}^{j}
\beta_{j}\beta^{k}(G^{\alpha})_{k}^{l}\beta_{l},
\end{equation}
where $d^{ijk}$, the symmetric invariant $E_{6}$ tensor, and
$(G^{\alpha})^{i}_{j}$ are defined in ref.\cite{Kephart}. From the
potential given in eq.(57) we can read directly the $F$-, $D$- and
scalar soft terms which break softly the supersymmetric theory
obtained by CSDR over $G_{2}/SU(3)$. The F-terms are obtained from
the superpotential
\begin{equation}
{\cal W} (B) =\frac{1}{3}d_{ijk}B^{i}B^{j}B^{k},
\end{equation}
where $B$ is the chiral superfield whose scalar component is
$\beta$. Let us note that the superpotential could also be
identified from the relevant Yukawa terms of the fermion part of
the Lagrangian. Correspondingly the $D$-terms are
\begin{equation}
D^{\alpha}
=\sqrt{\frac{11}{2}}\beta^{i}(G^{\alpha})^{j}_{i}\beta_{j}.
\end{equation}
The terms in the potential $V(\beta)$ given in eq.(57) that do not
result from the $F$- and $D$-terms belong to the soft
supersymmetry part of the Lagrangian. These terms are the
following,
\begin{equation}
{\cal L}_{scalarSSB} =
-\frac{40}{3R_{1}^{2}}\beta^{2}-\left[\frac{4}{R_{1}}d_{ijk}\beta^{i}\beta^{j}\beta^{k}+
h.c \right].
\end{equation}

Note that the potential (57) belongs to the case, discussed in
subsection $2.3$, that $S$ can be embedded in $G$ \cite{LZ}.
Finally in order to determine the gaugino mass we calculate the V
operator given in eq.(40). Using eq.(42) we find that
\begin{equation}
D_{abc}=\frac{R_{1}^{2}}{2}f_{abc}
\end{equation}
and in turn the $G_{abc}=D_{abc}+\frac{1}{2}T_{abc}$ is
\begin{equation}
G_{abc}=\frac{R_{1}^{2}}{2}(1+3\tau)f_{abc}.
\end{equation}
In order to obtain the previous results the most general $G_{2}$
invariant metric on $G_{2}/SU(3)$ was used which is
$g_{ab}=R_{1}^{2}\delta_{ab}$.\\ In addition we  need the gamma
matrices in ten dimensions given in appendix B. We find that the
gauginos acquire a geometrical mass
\begin{equation}
M=(1+3\tau)\frac{6}{\sqrt{3}R_{1}}.
\end{equation}

\vspace{.1in}

{\it {\bf b. Soft Supersymmerty breaking by dimensional reduction
over\\ $Sp(4)/(SU(2) \times U(1))_{non-max.}$ }} \\

In this case we start again with a ten-dimensional supersymmetric
gauge theory based on the
 group $E_{8}$ and reduce it over the non-symmetric coset
 $Sp(4)/(SU(2) \times U(1))_{non-max.}$. Therefore we have chosen
$G=E_{8}$ and $B=Sp(4)/(SU(2) \times U(1))_{non-max.}$. We begin
by giving the decompositions to be used, $$E_{8} \supset SU(3)
\times E_{6} \supset SU(2) \times U(1) \times E_{6}.$$ The
decomposition of $248$ of $E_{8}$ under $SU(3) \times E_{6}$ is
given in eq.(50) while under $ (SU(2) \times U(1)) \times E_{6}$
is the following,
\begin{eqnarray}
248 =
(3,1)_{0}+(1,1)_{0}+(1,78)_{0}+(2,1)_{3}+(2,1)_{-3}\nonumber\\
+(1,27)_{-2}
+(1,\overline{27})_{2}+(2,27)_{1}+(2,\overline{27})_{-1}.
\end{eqnarray}
In the present case $R$ is chosen to be identified with the $SU(2)
\times U(1)$ of the latter of the above decompositions. Therefore
the resulting four-dimensional gauge group is $H=C_{E_{8}}(SU(2)
\times U(1))= E_{6} \times U(1)$. The $R=SU(2) \times U(1)$
content of $Sp(4)/(SU(2) \times U(1))_{non-max.}$ vector and
spinor according to table 2 are $1_{2}+1_{-2}+2_{1}+2_{-1}$ and
$1_{0}+1_{-2}+2_{1}$ respectively. Thus applying the CSDR rules
(24),(25) and (26),(27) we find that the surviving fields in four
dimensions can be organized in a ${\cal N}=1$ vector
supermultiplet $V^{\alpha}$ which transforms as $78$ of $E_{6}$, a
${\cal N}=1$ $U(1)$ vector supermultiplet $V$ and two chiral
supermultiplets ($B^{i}$, $C^{i}$), transforming as $(27,1)$ and
$(27,-2)$ under $E_{6} \times U(1)$.

To determine the potential we have to go in the details and
examine further the decomposition of the adjoint of the specific
$S$ under $R$, i.e. $$Sp(4) \supset (SU(2) \times
U(1))_{non-max.}$$
\begin{equation}
10 = 3_{0}+1_{0}+1_{2}+1_{-2}+2_{1}+2_{-1}.
\end{equation}
Then, according to the decomposition (65) the generators of
$Sp(4)$ can be grouped as follows,
\begin{equation}
Q_{Sp(4)} = \{ Q^{\rho},Q,Q_{+},Q^{+},Q^{a},Q_{a}\},
\end{equation}
where $\rho$ takes values $1,2,3$ and $a$ takes the values $1,2$.
The non-trivial commutation relations of the $Sp(4)$ generators
given in (66) are given in table\ 5 of appendix A. Furthermore the
decomposition (66) suggests the following change in the notation
of the scalar fields
\begin{equation}
\{ \phi_{I}, I=1,\ldots,10\} \longrightarrow ( \phi^{\rho}, \phi,
\phi_{+}, \phi^{+}, \phi^{a}, \phi_{a}),
\end{equation}
which facilitates the solution of the constraints (33).

On the other hand the potential of any gauge theory reduced over
the coset space $Sp(4)/(SU(2) \times U(1))_{non-max.}$ was found
\cite{Review} to be in terms of the redefined fields in (67),
\begin{eqnarray} V(\phi) =
\frac{2\Lambda^{2}+6}{R_{1}^{4}}+\frac{4\Lambda^{2}}{R_{2}^{4}}+
\frac{2}{R_{2}^{4}}Tr(\phi_{+}\phi^{+})+
\frac{2}{R_{1}^{4}}Tr(\phi_{a}\phi^{a})\nonumber\\
-\frac{2\Lambda}{R_{2}^{4}}Tr(Q[\phi_{+},\phi^{+}])
-\frac{\Lambda}{R_{1}^{4}}Tr(Q[\phi_{a},\phi^{a}])
-\frac{1}{R_{1}^{4}}(\tau_{\rho})^{a}_{b}Tr(Q_{\rho}[\phi_{a},\phi^{b}])\nonumber\\
\biggl[-\frac{\sqrt{2}}{R_{1}^{2}}(\frac{1}{R_{2}^{2}}+\frac{1}{2R_{1}^{2}})\epsilon^{ab}
Tr(\phi_{+}[\phi_{a},\phi_{b}]) + h.c\biggr]\nonumber\\ +
\frac{1}{2}(\frac{1}{R_{2}^{2}}[\phi_{+},\phi^{+}]+
\frac{1}{R_{1}^{2}}[\phi_{a},\phi^{a}])^{2}\nonumber\\
-\frac{2}{R_{1}^{2}R_{2}^{2}}Tr([\phi_{+},\phi_{a}][\phi^{+},\phi^{a}])
-\frac{1}{R_{1}^{4}}Tr([\phi_{a},\phi_{b}][\phi^{a},\phi^{b}]),
\end{eqnarray}
where, $R_{1}$ and $R_{2}$ are the coset space radii. In terms of
the radii the real metric\footnote{The coset space can be
considered as a complex three-dimensional space having coordinate
indices $a,+$ with $a=1,2$ and metric
$g^{1\overline{1}}=g^{2\overline{2}}=\frac{1}{R_{1}^{2}}$ and
$g^{+\overline{+}}=\frac{1}{R_{2}^{2}}$. The latter metric has
been used to write the potential in the form given in eq.(50).}
of the coset space is
\begin{equation}
g_{ab}=diag(R_{1}^{2},R_{1}^{2},R_{2}^{2},R_{2}^{2},R_{1}^{2},R_{1}^{2})
\end{equation}

To proceed we use the embedding (64) of $SU(2) \times U(1)$ in
$E_{8}$ and divide its generators accordingly,
\begin{equation}
Q_{E_{8}} = \{ G^{\rho}, G, G^{\alpha}, G^{a}, G_{a}, G^{i},
G_{i}, G^{ai}, G_{ai} \}
\end{equation}
where, $\rho = 1,2,3$, $a=1,2$, $\alpha=1,\ldots,78$,
$i=1,\ldots,27$. The non-trivial commutation relations of the
$E_{8}$ generators grouped in (70) are given in table 6 of the
appendix A.

Now the constraints (33) for the redefined fields in (67) become
\begin{equation}
\left[\phi,\phi_{+}\right] = 2\phi_{+},\
\left[\phi,\phi_{a}\right]= \phi_{a},\
\left[\phi_{\rho},\phi_{a}\right]=(\tau_{\rho})^{b}_{a}\phi_{b}.
\end{equation}
Then we can write the solutions of the constraints  (71) in terms
of the genuine Higgs fields $\beta^{i}$, $\gamma^{i}$ and the
$E_{8}$ generators (70) corresponding to the embedding (64) as
follows,
\begin{eqnarray}
\phi^{\rho}=G^{\rho}, \phi=\sqrt{3}G, \nonumber \\
\phi_{a}=R_{1}\frac{1}{\sqrt{2}}\beta^{i}G_{1i},
\phi_{+}=R_{2}\gamma^{i}G_{i}.
\end{eqnarray}

The potential (68) in terms of the physical scalar fields
$\beta^{i}$ and $\gamma^{i}$ becomes
\begin{eqnarray}
V(\beta^{i},\gamma^{i})= const -\frac{6}{R_{1}^{2}}\beta^{2}
-\frac{4}{R_{2}^{2}}\gamma^{2} \nonumber\\
+\bigg[4\sqrt{\frac{10}{7}}R_{2}
\bigl(\frac{1}{R_{2}^{2}}+\frac{1}{2R_{1}^{2}}\bigr)
d_{ijk}\beta^{i}\beta^{j}\gamma^{k} + h.c \biggr] \nonumber \\
+6\biggl(\beta^{i}(G^{\alpha})_{i}^{j}\beta_{j}
+\gamma^{i}(G^{\alpha})_{i}^{j}\gamma_{j}\biggr)^{2}+\nonumber\\
\frac{1}{3}\biggl(\beta^{i}(1\delta_{i}^{j})\beta_{j}+
\gamma^{i}(-2\delta_{i}^{j})\gamma_{j}\biggr)^{2}\nonumber\\
+\frac{5}{7}\beta^{i}\beta^ {j}d_{ijk}d^{klm}\beta_{l}\beta_{m}
+4\frac{5}{7}\beta^{i}\gamma^{j}d_{ijk}d^{klm}\beta_{l}\gamma_{m}.
\end{eqnarray}
From the potential (73) we read the $F$-, $D$- and scalar soft
terms as in the previous model. The $F$-terms can be derived from
the superpotential
\begin{equation}
{\cal W}(B^{i},C^{j})= \sqrt{\frac{5}{7}}d_{ijk}B^{i}B^{j}C^{k}.
\end{equation}
The $D$-term contributions are the sum
\begin{equation}
\frac{1}{2}D^{\alpha}D^{\alpha}+\frac{1}{2}DD,
\end{equation}
where
$$D^{\alpha}=\sqrt{12}\bigl(\beta^{i}(G^{\alpha})_{i}^{j}\beta_{j}
+\gamma^{i}(G^{\alpha})_{i}^{j}\gamma_{j}\bigr)$$ and $$
D=\sqrt{\frac{2}{3}}\bigl(\beta^{i}(1\delta_{i}^{j})\beta_{j}+
\gamma^{i}(-2\delta_{i}^{j})\gamma_{j}\bigr)$$ corresponding to
$E_{6} \times U(1)$. The rest terms in the potential (73) are the
soft breaking mass and trilinear terms and they form the scalar
SSB part of the Lagrangian,
\begin{equation}
{ \cal L}_{scalarSSB}= -\frac{6}{R_{1}^{2}}\beta^{2}
-\frac{4}{R_{2}^{2}}\gamma^{2} + \bigg[4\sqrt{\frac{10}{7}}R_{2}
\bigl(\frac{1}{R_{2}^{2}}+\frac{1}{2R_{1}^{2}}\bigr)
d_{ijk}\beta^{i}\beta^{j}\gamma^{k} + h.c \biggr].
\end{equation}
 The gaugino mass has been calculated in ref.\cite{Kapetanakis} to be
\begin{equation}
M=(1+3\tau)\frac{R_{2}^{2}+2R_{1}^{2}}{8R_{1}^{2}R_{2}}.
\end{equation}
We note that the chosen embedding of $R=SU(2) \times U(1)$ in
$E_{8}$ satisfies the condition (30) which guarantees the
renormalizability of the four-dimensional theory, while the
absence of any other term that does not belong to the
supersymmetric $E_{6} \times U(1)$ theory or to its SSB sector
guarantees the improved ultraviolet behaviour of the theory as in
the previous model. Finally note the contribution of the torsion
in the gaugino mass (77).\\

\vspace{.2in}

{\it {\bf c. Soft Supersymmetry breaking by reduction over
$SU(3)/(U(1) \times U(1))$.}}\\

In this model the only difference as compared to the previous ones
is that the chosen coset space to reduce the same theory is the
non-symmetric $B=SU(3)/U(1) \times U(1)$. The decompositions to be
used are $$ E_{8} \supset SU(2) \times U(1) \times E_{6} \supset
U(1) \times U(1) \times E_{6} $$ The $248$ of $E_{8}$ is
decomposed under $SU(2) \times U(1)$ according to (64) whereas the
decomposition under $U(1) \times U(1)$ is the following:
\begin{eqnarray}
 248 = 1_{(0,0)}+1_{(0,0)}+1_{(3,\frac{1}{2})}+1_{(-3,\frac{1}{2})}+\nonumber\\
1_{(0,-1)}+1_{(0,1)}+1_{(-3,-\frac{1}{2})}+1_{(3,-\frac{1}{2})}+\nonumber\\
78_{(0,0)}+27_{(3,\frac{1}{2})}+27_{(-3,\frac{1}{2})}+27_{(0,-1)}+\nonumber\\
\overline{27}_{(-3,-\frac{1}{2})}+\overline{27}_{(3,-\frac{1}{2})}
+\overline{27}_{(0,1)}.
\end{eqnarray}
In the present case $R$ is chosen to be identified with the $U(1)
\times U(1)$ of the latter decomposition. Therefore the resulting
four-dimensional gauge group is $$ H=C_{E_{8}}(U(1) \times U(1)) =
 U(1) \times U(1) \times E_{6} $$  The $R=U(1) \times U(1)$
content of $SU(3)/U(1) \times U(1)$ vector and spinor are
according to table 2, $$(3,\frac{1}{2})+(-3,\frac{1}{2})
+(0,-1)+(-3,-\frac{1}{2})+(3,-\frac{1}{2})+(0,1)$$ and
$$(0,0)+(3,\frac{1}{2})+(-3,\frac{1}{2}) +(0,-1)$$ respectively.
Thus applying the CSDR rules we find that the surviving fields in
four dimensions are three ${\cal N}=1$ vector multiplets
$V^{\alpha},V_{(1)},V_{(2)}$, (where $\alpha$ is an $E_{6}$, $78$
index and the other two refer to the two $U(1)'s$) containing the
gauge fields of $U(1) \times U(1) \times E_{6}$. The matter
content consists of three ${\cal N}=1$ chiral multiplets ($A^{i}$,
$B^{i}$, $C^{i}$) with $i$ an $E_{6}$, $27$ index and three ${\cal
N}=1$ chiral multiplets ($A$, $B$, $C$) which are $E_{6}$ singlets
and carry $U(1) \times U(1)$ charges.

To determine the potential we examine further the decomposition
of the adjoint of the specific $S=SU(3)$ under $R=U(1) \times
U(1)$, i.e.
$$SU(3) \supset U(1) \times U(1) $$
\begin{eqnarray}
 8 = (0,0)+(0,0)+(3,\frac{1}{2})+(-3,\frac{1}{2})
+(0,-1)+\nonumber\\(-3,-\frac{1}{2})+(3,-\frac{1}{2})+(0,1).
\end{eqnarray}
Then according to the decomposition (79) the generators of $SU(3)$
can be grouped as
\begin{equation}
Q_{SU(3)} = \{Q_{0},Q'_{0},Q_{1},Q_{2},Q_{3},Q^{1},Q^{2},Q^{3} \}.
\end{equation}
The non trivial commutator relations of $SU(3)$ generators (80)
are given in table 7 of the appendix A. The decomposition (80)
suggests the following change in  the notation of the scalar
fields,
\begin{equation}
(\phi_{I}, I=1,\ldots,8) \longrightarrow ( \phi_{0}, \phi'_{0},
\phi_{1}, \phi^{1}, \phi_{2}, \phi^{2}, \phi_{3}, \phi^{3}).
\end{equation}

The potential of any theory reduced over $SU(3)/U(1) \times U(1))$
is given in terms of the redefined fields in (81) by
\begin{eqnarray}
\lefteqn{V(\phi)=(3\Lambda^{2}+\Lambda'^{2})\biggl(\frac{1}{R_{1}^{4}}+\frac{1}{R_{2}^{4}}\biggr)
+\frac{4\Lambda'^{2}}{R_{3}^{2}}}\nonumber\\
&&+\frac{2}{R_{2}^{2}R_{3}^{2}}Tr(\phi_{1}\phi^{1})+
\frac{2}{R_{1}^{2}R_{3}^{2}}Tr(\phi_{2}\phi^{2})
+\frac{2}{R_{1}^{2}R_{2}^{2}}Tr(\phi_{3}\phi^{3})\nonumber\\
&&+\frac{\sqrt{3}\Lambda}{R_{1}^{4}}Tr(Q_{0}[\phi_{1},\phi^{1}])
-\frac{\sqrt{3}\Lambda}{R_{2}^{4}}Tr(Q_{0}[\phi_{2},\phi^{2}])
-\frac{\sqrt{3}\Lambda}{R_{3}^{4}}Tr(Q_{0}[\phi_{3},\phi^{3}])\nonumber\\
&&+\frac{\Lambda'}{R_{1}^{4}}Tr(Q'_{0}[\phi_{1},\phi^{1}])
+\frac{\Lambda'}{R_{2}^{4}}Tr(Q'_{0}[\phi_{2},\phi^{2}])
-\frac{2\Lambda'}{R_{3}^{4}}Tr(Q'_{0}[\phi_{3},\phi^{3}])\nonumber\\
&&+\biggl[\frac{2\sqrt{2}}{R_{1}^{2}R_{2}^{2}}Tr(\phi_{3}[\phi_{1},\phi_{2}])
+\frac{2\sqrt{2}}{R_{1}^{2}R_{3}^{3}}Tr(\phi_{2}[\phi_{3},\phi_{1}])
+\frac{2\sqrt{2}}{R_{2}^{2}R_{3}^{2}}Tr(\phi_{1}[\phi_{2},\phi_{3}])+ h.c\biggr]\nonumber\\
&&+\frac{1}{2}Tr \biggl(\frac{1}{R_{1}^{2}}([\phi_{1},\phi^{1}])+
\frac{1}{R_{2}^{2}}([\phi_{2},\phi^{2}])
+\frac{1}{R_{3}^{2}}([\phi_{3},\phi^{3}])\biggr)^{2}\nonumber\\
&&-\frac{1}{R_{1}^{2}R_{2}^{2}}Tr([\phi_{1},\phi_{2}][\phi^{1},\phi^{2}])
-\frac{1}{R_{1}^{2}R_{3}^{2}}Tr([\phi_{1},\phi_{3}][\phi^{1},\phi^{3}])\nonumber\\
&&-\frac{1}{R_{2}^{2}R_{3}^{2}}Tr([\phi_{2},\phi_{3}][\phi^{2},\phi^{3}]),
\end{eqnarray}
where $R_{1},R_{2},R_{3}$ are the coset space radii\footnote{To
bring the potential into this form we have used (A.22) of
ref.\cite{Review} and relations (7),(8) of ref.\cite{Witten2}.}.
In terms of the radii the real metric\footnote{The complex metric
that was used is
$g^{1\overline{1}}=\frac{1}{R_{1}^{2}},g^{2\overline{2}}=\frac{1}{R_{2}^{2}},
g^{3\overline{3}}=\frac{1}{R_{3}^{2}}$.} of the coset is
\begin{equation}
g_{ab}=diag(R_{1}^{2},R_{1}^{2},R_{2}^{2},R_{2}^{2},R_{3}^{2},R_{3}^{2}).
\end{equation}

Next we examine the commutation relations of $E_{8}$ under the
decomposition (78). Under this decomposition the generators of
$E_{8}$ can be grouped as
\begin{eqnarray}
Q_{E_{8}}=\{Q_{0},Q'_{0},Q_{1},Q_{2},Q_{3},Q^{1},Q^{2},Q^{3},Q^{\alpha},\nonumber\\
Q_{1i},Q_{2i},Q_{3i},Q^{1i},Q^{2i},Q^{3i} \},
\end{eqnarray}
where, $ \alpha=1,\ldots,78 $ and $ i=1,\ldots,27 $. The
non-trivial commutation relations of the $E_{8}$ generators (84)
are given in tables 8.1 and 8.2 of appendix A. \\ Now the
constraints (33) for the redefined fields in (81) are,
\begin{eqnarray}
\left[\phi_{1},\phi_{0}\right]=\sqrt{3}\phi_{1}&,&
\left[\phi_{1},\phi_{0}'\right]=\phi_{1}, \nonumber \\
\left[\phi_{2},\phi_{0}\right]=-\sqrt{3}\phi_{2}&,&
\left[\phi_{2},\phi_{0}'\right]=\phi_{2}, \nonumber \\
\left[\phi_{3},\phi_{0}\right]=0&,&
\left[\phi_{3},\phi_{0}'\right]=-2\phi_{3}.
\end{eqnarray}
The solutions of the constraints (85) in terms of the genuine
Higgs fields and of the $E_{8}$ generators (84) corresponding to
the embedding (78) of $R=U(1) \times U(1)$ in  the $E_{8}$ are,
$\phi_{0}=\Lambda Q_{0}$ and $\phi_{0}'=\Lambda Q_{0}'$,with
$\Lambda=\Lambda'=\frac{1}{\sqrt{10}}$, and
\begin{eqnarray}
\phi_{1} &=& R_{1} \alpha^{i} Q_{1i}+R_{1} \alpha Q_{1},
\nonumber\\ \phi_{2} &=& R_{2} \beta^{i} Q_{2i}+ R_{2} \beta
Q_{2}, \nonumber\\ \phi_{3} &=& R_{3} \gamma^{i} Q_{3i}+ R_{3}
\gamma Q_{3},
\end{eqnarray}
where the unconstrained  scalar fields transform under $U(1)
\times U(1) \times E_{6}$ as
\begin{eqnarray}
\alpha_{i} \sim 27_{(3,\frac{1}{2})}&,&\alpha \sim
1_{(3,\frac{1}{2})},\nonumber\\ \beta_{i} \sim
27_{(-3,\frac{1}{2})}&,&\beta \sim
1_{(-3,\frac{1}{2})},\nonumber\\ \gamma_{i} \sim
27_{(0,-1)}&,&\gamma \sim 1_{(0,-1)}.
\end{eqnarray}
The potential (82) becomes
\begin{eqnarray}
V(\alpha^{i},\alpha,\beta^{i},\beta,\gamma^{i},\gamma)= const. +
\biggl( \frac{4R_{1}^{2}}{R_{2}^{2}R_{3}^{2}}-\frac{8}{R_{1}^{2}}
\biggr)\alpha^{i}\alpha_{i} +\biggl(
\frac{4R_{1}^{2}}{R_{2}^{2}R_{3}^{2}}-\frac{8}{R_{1}^{2}}
\biggr)\overline{\alpha}\alpha \nonumber \\
+\biggl(\frac{4R_{2}^{2}}{R_{1}^{2}R_{3}^{2}}-\frac{8}{R_{2}^{2}}\biggr)
\beta^{i}\beta_{i}
+\biggl(\frac{4R_{2}^{2}}{R_{1}^{2}R_{3}^{2}}-\frac{8}{R_{2}^{2}}\biggr)
\overline{\beta}\beta \nonumber \\
+\biggl(\frac{4R_{3}^{2}}{R_{1}^{2}R_{2}^{2}}
-\frac{8}{R_{3}^{2}}\biggr)\gamma^{i}\gamma_{i}
+\biggl(\frac{4R_{3}^{2}}{R_{1}^{2}R_{2}^{2}}
-\frac{8}{R_{3}^{2}}\biggr)\overline{\gamma}\gamma \nonumber\\
+\biggl[\sqrt{2}80\biggl(\frac{R_{1}}{R_{2}R_{3}}+\frac{R_{2}}{R_{1}
R_{3}}+\frac{R_{3}}{R_{2}R_{1}}\biggr)d_{ijk}\alpha^{i}\beta^{j}\gamma^{k}\nonumber\\
+\sqrt{2}80\biggl(\frac{R_{1}}{R_{2}R_{3}}+\frac{R_{2}}{R_{1}
R_{3}}+\frac{R_{3}}{R_{2}R_{1}}\biggr)\alpha\beta\gamma+
h.c\biggr]\nonumber\\
+\frac{1}{6}\biggl(\alpha^{i}(G^{\alpha})_{i}^{j}\alpha_{j}
+\beta^{i}(G^{\alpha})_{i}^{j}\beta_{j}
+\gamma^{i}(G^{\alpha})_{i}^{j}\gamma_{j}\biggr)^{2}\nonumber\\
+\frac{10}{6}\biggl(\alpha^{i}(3\delta_{i}^{j})\alpha_{j} +
\overline{\alpha}(3)\alpha + \beta^{i}(-3\delta_{i}^{j})\beta_{j}
+ \overline{\beta}(-3)\beta \biggr)^{2}\nonumber \\
+\frac{40}{6}\biggl(\alpha^{i}(\frac{1}{2}\delta_{i}^{j})\alpha_{j}
+ \overline{\alpha}(\frac{1}{2})\alpha +
\beta^{i}(\frac{1}{2}\delta^{j}_{i})\beta_{j} +
\overline{\beta}(\frac{1}{2})\beta +
\gamma^{i}(-1\delta_{i}^{j})\gamma_{j} +
\overline{\gamma}(-1)\gamma \biggr)^{2}\nonumber \\
+40\alpha^{i}\beta^{j}d_{ijk}d^{klm}\alpha_{l}\beta_{m}
+40\beta^{i}\gamma^{j}d_{ijk}d^{klm}\beta_{l}\gamma_{m}
+40\alpha^{i}\gamma^{j}d_{ijk}d^{klm}\alpha_{l}\gamma_{m}\nonumber\\
+40(\overline{\alpha}\overline{\beta})(\alpha\beta) +
40(\overline{\beta}\overline{\gamma})(\beta\gamma) +
40(\overline{\gamma}\overline{\alpha})(\gamma\alpha).
\end{eqnarray}
From the potential (88) we read the $F$-, $D$- and scalar soft
terms. The $F$-terms are obtained from the superpotential
\begin{equation}
{ \cal W }(A^{i},B^{j},C^{k},A,B,C)
=\sqrt{40}d_{ijk}A^{i}B^{j}C^{k} + \sqrt{40}ABC.
\end{equation}
The $D$-terms have the structure
\begin{equation}
\frac{1}{2}D^{\alpha}D^{\alpha}+\frac{1}{2}D_{1}D_{1}+\frac{1}{2}D_{2}D_{2},
\end{equation}
where $$D^{\alpha}= \frac{1}{\sqrt{3}}
\biggl(\alpha^{i}(G^{\alpha})_{i}^{j}\alpha_{j}
+\beta^{i}(G^{\alpha})_{i}^{j}\beta_{j}
+\gamma^{i}(G^{\alpha})_{i}^{j}\gamma_{j}\biggr),$$ $$D_{1}=
\sqrt{ \frac{10}{3} }\biggl(\alpha^{i}(3\delta_{i}^{j})\alpha_{j}
+ \overline{\alpha}(3)\alpha +
\beta^{i}(-3\delta_{i}^{j})\beta_{j} + \overline{\beta}(-3)\beta
\biggr)$$ and $$D_{2} = \sqrt{ \frac{40}{3}
}\biggl(\alpha^{i}(\frac{1}{2}\delta_{i}^{j})\alpha_{j} +
\overline{\alpha}(\frac{1}{2})\alpha +
\beta^{i}(\frac{1}{2}\delta^{j}_{i})\beta_{j} +
\overline{\beta}(\frac{1}{2})\beta +
\gamma^{i}(-1\delta_{i}^{j})\gamma_{j} +
\overline{\gamma}(-1)\gamma \biggr),$$ which correspond to the
$E_{6} \times U(1)_{1} \times U(1)_{2}$ structure of the gauge
group. The rest terms are the trilinear and mass terms which break
supersymmetry softly and they form the scalar SSB part of the
Lagrangian,
\begin{eqnarray}
\lefteqn{{\cal L}_{scalarSSB}=  \biggl(
\frac{4R_{1}^{2}}{R_{2}^{2}R_{3}^{2}}-\frac{8}{R_{1}^{2}}
\biggr)\alpha^{i}\alpha_{i} +\biggl(
\frac{4R_{1}^{2}}{R_{2}^{2}R_{3}^{2}}-\frac{8}{R_{1}^{2}}
\biggr)\overline{\alpha}\alpha} \nonumber\\
& &
+\biggl(\frac{4R_{2}^{2}}{R_{1}^{2}R_{3}^{2}}-\frac{8}{R_{2}^{2}}\biggr)
\beta^{i}\beta_{i}
+\biggl(\frac{4R_{2}^{2}}{R_{1}^{2}R_{3}^{2}}-\frac{8}{R_{2}^{2}}\biggr)
\overline{\beta}\beta
+\biggl(\frac{4R_{3}^{2}}{R_{1}^{2}R_{2}^{2}}
-\frac{8}{R_{3}^{2}}\biggr)\gamma^{i}\gamma_{i}
+\biggl(\frac{4R_{3}^{2}}{R_{1}^{2}R_{2}^{2}}
-\frac{8}{R_{3}^{2}}\biggr)\overline{\gamma}\gamma \nonumber\\
& &
+\biggl[\sqrt{2}80\biggl(\frac{R_{1}}{R_{2}R_{3}}+\frac{R_{2}}{R_{1}
R_{3}}+\frac{R_{3}}{R_{2}R_{1}}\biggr)d_{ijk}\alpha^{i}\beta^{j}\gamma^{k}
\nonumber \\
& &+\sqrt{2}80\biggl(\frac{R_{1}}{R_{2}R_{3}}+\frac{R_{2}}{R_{1}
R_{3}}+\frac{R_{3}}{R_{2}R_{1}}\biggr)\alpha\beta\gamma+
h.c\biggr].
\end{eqnarray}

Note that the potential (88) belongs to the case analyzed in
subsection~$2.3$ where $S$ has an image in $G$. Here $S=SU(3)$ has
an image in $G=E_{8}$ \cite{LZ} so we conclude that the minimum of
the potential is zero. Finally in order to determine the gaugino
mass, we calculate the V operator using appendix B. We find that
the gauginos acquire a geometrical mass
\begin{equation}
M=(1+3\tau)\frac{(R_{1}^{2}+R_{2}^{2}+R_{3}^{2})}{8\sqrt{R_{1}^{2}R_{2}^{2}R_{3}^{2}}}.
\end{equation}
Note again that the chosen embedding satisfies the condition (30)
and the absence in the four-dimensional theory of any other term
that does not belong to the supersymmetric $E_{6} \times U(1)
\times U(1)$ gauge theory or to its SSB sector. The gaugino mass
(92), as in the two previous models, has a contribution from the
torsion of the coset space. A final remark concerning the gaugino
masses in all three models reduced over six-dimensional
non-symmetric coset spaces with torsion is that the adjustments
required to obtain the {\it canonical connection} lead also to
vanishing gaugino masses. Contrary to the gaugino mass term the
soft scalar terms of the SSB do not receive contributions from the
torsion. This is due to the fact that gauge fields, contrary to
fermions, do not couple to torsion.

\vspace{.1in}

 Concluding the present subsection, we would like to note that the fact that, starting with a ${\cal
N} = 1$ supersymmetric theory in ten dimensions, the CSDR leads to
the field content of an ${\cal N} = 1$ supersymmetric theory in
the case that the six-dimensional coset spaces used are
non-symmetric, can been seen by inspecting the table 2. More
specifically, one notices in table $2$ that when the coset spaces
are non-symmetric the decompositions of the spinor $4$ and
antispinor $\overline{4}$ of $SO(6)$ under $R$ contain a singlet,
i.e. have the form $1+r$ and $1+\overline{r}$, respectively, where
$r$ is possibly reducible. The singlet under $R$ provides the
four-dimensional theory with fermions transforming according to
the adjoint as was emphasized in subsection $2.3$ and correspond
to gauginos, which obtain geometrical and torsion mass
contributions as we have seen in all three cases of the present
subsection $3.2$. Next turning the decomposition of the vector $6$
of $SO(6)$ under $R$ in the non-symmetric cases, we recall that
the vector can be constructed from the tensor product $4 \times 4$
and therefore has the form $r+\overline{r}$. Then the CSDR
constraints tell us that the four-dimensional theory will contain
the same representations of fermions and scalars since both come
from the adjoint representation of the gauge group $G$ and they
have to satisfy the same matching conditions under $R$. Therefore
the field content of the four-dimensional theory is, as expected,
${\cal N} =1$ supersymmetric. To find out that furthermore the
${\cal N} =1$ supersymmetry is softly broken, requires the lengthy
and detailed analysis that was done above.
\section{Conclusions}
The CSDR was originally introduced as a scheme which, making use
 of higher dimensions, incorporates in a
unified manner the gauge and the ad-hoc Higgs sector of the
spontaneously broken gauge theories in four dimensions
\cite{Manton}. Next fermions were introduced in the scheme and the
ad-hoc Yukawa interactions have also been included in the unified
description \cite{Slansky,Chapline}.

Of particular interest for the construction of fully realistic
theories in the framework of CSDR are the following virtues that
complemented the original suggestion: (i) The possibility to
obtain chiral fermions in four dimensions resulting from
vector-like representations of the higher dimensional gauge theory
\cite{Chapline,Review}. This possibility can be realized due the
presence of non-trivial background gauge configurations which are
introduced by the CSDR constructions \cite{Salam}, (ii) The
possibility to deform the metric of certain non-symmetric coset
spaces and thereby obtain more than one scales
\cite{Farakos,Review,Hanlon}, (iii) The possibility to use coset
spaces, which are multiply connected. This can be achieved by
exploiting the discrete symmetries of the S/R
\cite{Kozimirov,Review}. Then one might introduce topologically
non-trivial gauge field \cite{Zoupanos} configurations with
vanishing field strength and induce additional breaking of the
gauge symmetry. It is the Hosotani mechanism \cite{Hosotani}
applied in the CSDR.

In the above list recently has been added the interesting
possibility that the popular softly broken supersymmetric
four-dimensional chiral gauge theories might have their origin in
a higher dimensional supersymmetric theory with only vector
supermultiplet \cite{Pman}, which is dimensionally reduced over
non-symmetric coset spaces.

In the present paper we have presented explicit and detailed
examples of CSDR of a supersymmetric $E_{8}$ gauge theory over all
possible six-dimensional coset spaces. Out of our study there are
two cases that single out for further study as candidates to
describe realistically the observed low energy world. Both are
known GUTs containing three fermion families and scalars
appropriate for the spontaneous electroweak breaking. One case is
based on the reduction of the $E_{8}$ over the symmetric coset
space $SU(3) \times SU(2)/SU(2) \times U(1) \times U(1)$ and leads
to an $SO(10)$-type non supersymmetric GUT in four dimensions. The
other is based of the reduction of the same ten-dimensional gauge
group over the non-symmetric coset space $SU(3)/U(1)\times U(1)$
and leads to an $E_{6}$-type softly broken supersymmetric GUT in
four dimensions. Both require some additional mechanism to break
the four-dimensional GUT gauge group. Such a possibility is
offered by the Hosotani mechanism, mentioned already, and in both
cases there exist discrete symmetries acting freely on the
corresponding coset spaces that can been used. We plan to return
with a complete analysis of the possibilities to extract viable
phenomenology from both models.

The current discussion on the higher dimensional theories with
large extra dimensions provides a new framework to examine further
the CSDR. An obvious advantage is a reexamination of CSDR over
symmetric coset spaces. The fact that the four-dimensional scalar
potential obtained from the reduction over symmetric coset spaces
is tachyonic and appropriate for the electroweak symmetry breaking
excludes the possibility of radii with size of the order of
inverse of GUT or Planck scales, contrary to radii of inverse TeV
scale. Similarly it is worth reexamining the cases that $S$ can be
embedded in the higher dimensional gauge group $G$ and therefore
the final gauge group after spontaneous symmetry breaking can be
determined group theoretically. Again the spontaneous symmetry
breaking is appropriate for the electroweak symmetry breaking,
while there are not known examples that such a breaking is
suitable for the GUT breaking. These latter cases provide also the
advantage that the resulting four-dimensional theory has vanishing
cosmological constant. Finally the classical treatment used in
CSDR is justified in the case of large radii which are far away
from the scales that the quantum effects of gravity are
important.\\
\section*{Acknowledgements}
We would like to thank L.~Alvarez-Gaume, C.~Bachas, G.~L.~Cardoso,
P.~Forgacs, A.~Kehagias, C.~Kounnas, G.~Koutsoumbas, D.~Luest and
D.~Suematsu for useful discussions.
\newpage

{\Large {\bf Appendix A.} } \\

In this appendix we collect the tables of the six-dimensional
coset spaces $S/R$ with S simple or semisimple and $rankS=rankR$,
and the tables of the commutation relations needed for our
calculations. \\ \vspace{.2in}

\begin{centering}
\begin{tabular}{|l l l|} \hline
\multicolumn{3}{|c|}{Table 1}\\ \multicolumn{3}{|c|}{
Six-dimensional symmetric cosets with $rankS=rankR$}\\ \hline
$S/R$ & $SO(6)$ vector & $SO(6)$ spinor \\ \hline $SO(7)/SO(6)$ &
$6$ & $4$ \\ $SU(4)/SU(3) \times U(1)$ & $3_{-2}+\overline{3}_{2}$
& $1_{3}+3_{-1}$ \\  $Sp(4)/(SU(2) \times U(1))_{max}$ &
$3_{-2}+3_{2}$ & $1_{3}+3_{-1}$ \\ $SU(3) \times SU(2)/SU(2)
\times U(1) \times U(1)$ & $1_{0,2a}+1_{0,-2a}$ &
$1_{b,-a}+1_{-b,-a}$ \\  & $+2_{b,0}+2_{-b,0}$ & $+2_{0,a}$ \\
$Sp(4) \times SU(2)/SU(2) \times SU(2) \times U(1)$ &
$(1,1)_{2}+(1,1)_{-2}$ & $(2,1)_{1}+(1,2)_{-1}$ \\  & $+(2,2)_{0}$
& \\ $(SU(3)/U(1))^{3}$ & $(2a,0,0)+(-2a,0,0)$ &
$(a,b,c)+(-a,-b,c)$ \\  & $+(0,2b,0)+(0,-2b,0)$  &
$+(-a,b,-c)+(a,-b,-c)$ \\ & $(0,0,2c)+(0,0,-2c)$ & \\ \hline
\end{tabular} \\ \vspace{.2in}

\begin{tabular}{|l l l|} \hline
\multicolumn{3}{|c|}{Table 2}\\ \multicolumn{3}{|c|}{
Six-dimensional non-symmetric cosets with $rankS=rankR$}\\ \hline
$S/R$ & $SO(6)$ vector & $SO(6)$ spinor \\ \hline $G_{2}/SU(3)$ &
$3+\overline{3}$ & $1+3$ \\ $Sp(4)/(SU(2) \times U(1))_{non-max}$
& $1_{2}+1_{-2}+2_{1}+2_{-1}$ & $1_{0}+1_{2}+2_{-1}$ \\
$SU(3)/U(1) \times U(1)$ & $(a,c)+(b,d)+(a+b,c+d)$ &
$(0,0)+(a,c)+(b,d)$ \\ & $+(-a,-c)+(-b,-d)$ & $+(-a-b,-c-d)$ \\ &
$+(-a-b,-c-d)$ &  \\ \hline
\end{tabular} \\ \vspace{.2in}

\begin{tabular}{|l|l|}\hline
\multicolumn{2}{|c|}{Table 3}\\ \multicolumn{2}{|c|}{Non-trivial
commutation relations of $G_{2}$ according to}\\
\multicolumn{2}{|c|}{the decomposition given in eq.(52)}\\ \hline
$\left[ Q^{a},Q^{b} \right] = 2i f^{abc}Q^{c}$ & $\left[
Q^{a},Q^{\rho} \right] = -(\lambda^{a})^{\rho}_{\sigma}Q^{\sigma}$
\\
$\left[ Q^{\rho},Q_{\sigma} \right] =
-(\lambda^{a})^{\rho}_{\sigma}Q^{a}$ & $\left[ Q^{\rho},Q^{\sigma}
\right] = 2\sqrt{\frac{2}{3}}\epsilon^{\rho\sigma\tau}Q_{\tau}$
\\ \hline
\end{tabular} \\ \vspace{.2in}

The normalization is $$TrQ^{a}Q^{b}=2\delta^{ab},
TrQ^{\rho}Q_{\sigma}=2\delta^{\rho}_{\sigma} . $$

\begin{tabular}{|l|l|}\hline
\multicolumn{2}{|c|}{Table 4}\\ \multicolumn{2}{|c|}{Non-trivial
commutation relations of $E_{8}$ according to} \\
\multicolumn{2}{|c|} {the decomposition given in eq.(55)}
\\ \hline
$\left[ Q^{a},Q^{b} \right]=2if^{abc}Q^{c}$ & $\left[
Q^{\alpha},Q^{\beta}
\right]=2ig^{\alpha\beta\gamma}Q^{\gamma}$ \\
$\left[ Q^{a},Q^{i\rho}
\right]=-(\lambda^{\alpha})^{\rho}_{\sigma}\delta^{i}_{j}Q^{j\sigma}$
& $\left[ Q^{i\rho},Q^{j\sigma} \right]=\frac{1}{\sqrt{6}}
\epsilon^{\rho\sigma\tau}d^{ijk}Q_{k\tau}$ \\
$\left[ Q^{i\rho},Q_{j\sigma}
\right]=-(\lambda^{a})^{\rho}_{\sigma}\delta^{i}_{j}Q^{a}+$ & $
\left[ Q^{\alpha},Q^{i\rho}
\right]=(G^{\alpha})^{i}_{j}\delta^{\rho}_{\sigma}Q^{j\sigma}$ \\
$\frac{1}{6}\delta^{\rho}_{\sigma}(G^{\alpha})^{i}_{j}Q^{\alpha}$
& \\ \hline
\end{tabular} \\ \vspace{.2in}

The normalization is $$TrQ^{a}Q^{b}=2\delta^{ab},\
TrQ^{\alpha}Q^{\beta}=12\delta^{\alpha\beta},\
TrQ^{i\rho}Q_{j\sigma}=2\delta^{i}_{j}\delta^{\rho}_{\sigma}.$$

\begin{tabular}{|l|l|l|} \hline
\multicolumn{3}{|c|}{Table 5}\\ \multicolumn{3}{|c|}{Non-trivial
commutation relations of $Sp(4)$ according to}\\
\multicolumn{3}{|c|}{the decomposition given in eq.(66)}\\ \hline
$\left[Q_{\rho},Q_{\sigma}\right]=
2i\epsilon_{\rho\sigma\tau}Q_{\tau}$ & $\left[Q,Q_{a}\right] =
Q_{a}$ & $\left[Q_{\rho},Q_{a}\right] =
(\tau_{\rho})^{b}_{a}Q_{b}$ \\ $\left[Q,Q_{+}\right] = 2Q_{+}$ &
$\left[Q_{a},Q^{+}\right] = -\sqrt{2}\epsilon_{ab}Q^{b}$ &
$\left[Q_{a},Q_{b}\right] = \sqrt{2}\epsilon_{ab}Q_{+}$ \\
$\left[Q_{a},Q^{b}\right] =
\delta^{a}_{b}Q+(\tau_{\rho})^{b}_{a}Q_{\rho}$ &
$\left[Q_{+},Q^{+}\right] = 2Q$ &  \\ \hline
\end{tabular} \\ \vspace{.2in}

The normalization in the above table is given by\\
$Tr(Q_{\rho}Q_{\sigma})=2\delta_{\rho\sigma}$,
$Tr(Q_{a}Q^{b})=2\delta^{b}_{a}$, $Tr(Q_{+}Q^{+})=2$. \\
\vspace{.2in}

\begin{tabular}{|l|l|} \hline
\multicolumn{2}{|c|}{Table 6}\\  \multicolumn{2}{|c|}{Non-trivial
commutation relations of $E_{8}$ according to} \\
\multicolumn{2}{|c|} {the decomposition given in eq.(70)}
\\ \hline
$\left[G^{\alpha},G^{\beta}\right] =
2ig^{\alpha\beta\gamma}G^{\gamma}$ &
$\left[G^{\rho},G^{\sigma}\right] =
2i\epsilon^{\rho\sigma\tau}G^{\tau}$ \\ $\left[G,G^{a}\right] =
\sqrt{3}G^{a}$ & $\left[G,G^{j}\right] = -\frac{2}{\sqrt{3}}G^{j}$
\\ $\left[G,G^{aj}\right] = \frac{1}{\sqrt{3}}G^{aj}$ &
$\left[G^{\rho},G^{a}\right] = -(\tau^{\rho})^{a}_{b}G^{b}$ \\
$\left[G^{\alpha},G^{i}\right] = -(G^{\alpha})^{i}_{j}G^{j}$ &
$\left[G^{\alpha},G^{ai}\right] = -(G^{\alpha})^{i}_{j}G^{aj}$ \\
$\left[G^{a},G^{j}\right] = \sqrt{2}G^{aj}$ &
$\left[G_{a},G^{bj}\right] = \sqrt{2}\delta^{b}_{a}G^{j}$ \\
$\left[G^{i},G^{aj}\right] =
-\sqrt{\frac{5}{7}}\epsilon^{ab}d^{ijk}G_{bk}$ &
$\left[G^{ai},G^{bj}\right] =
\sqrt{\frac{5}{7}}\epsilon^{ab}d^{ijk}G_{k}$ \\
$\left[G^{i},G_{aj}\right] = \sqrt{2}\delta^{i}_{j}G_{a}$ &
$\left[G^{a},G_{b}\right] =
\sqrt{3}\delta^{a}_{b}-(\tau^{\rho})^{a}_{b}G^{\rho}$ \\
$\left[G^{i},G_{j}\right] = -\frac{2}{\sqrt{3}}\delta^{i}_{j}G
+(G^{\alpha})^{i}_{j}G^{\alpha}$ & $\left[G^{ai},G_{bj}\right] =
\frac{1}{\sqrt{3}}\delta^{i}_{j}\delta^{a}_{b}G
+\delta^{a}_{b}(G^{\alpha})^{i}_{j}G^{\alpha}
-\delta^{i}_{j}(\tau^{\rho})^{a}_{b}G^{\rho}$ \\ \hline
\end{tabular} \\ \vspace{.2in}

The normalization in the above table is as follows \\
$$Tr(G^{\rho}G_{\sigma})=2\delta^{\rho\sigma},
Tr(G^{\alpha}G^{\beta})=12\delta^{\alpha\beta},
Tr(G^{a}G_{b})=2\delta^{a}_{b}$$ $$Tr(GG)=2,
Tr(G^{i}G_{j})=2\delta^{i}_{j},
Tr(G^{ai}G_{bj})=2\delta^{a}_{b}\delta^{i}_{j}$$. \\ \vspace{.2in}

\begin{tabular}{|l|l|l|}\hline
\multicolumn{3}{|c|}{Table 7}\\ \multicolumn{3}{|c|}{Non-trivial
commutation relations of $SU(3)$ according to} \\
\multicolumn{3}{|c|}{the decomposition given in eq.(80)}\\ \hline
$\left[Q_{1},Q_{0}\right]=\sqrt{3}Q_{1}$ &
$\left[Q_{1},Q'_{0}\right]=Q_{1}$ &
$\left[Q_{2},Q_{0}\right]=-\sqrt{3}Q_{2}$\\
$\left[Q_{2},Q'_{0}\right]=Q_{2}$ & $\left[Q_{3},Q_{0}\right]=0$ &
$\left[Q_{3},Q'_{0}\right]=-2Q_{3}$ \\
$\left[Q_{1},Q^{1}\right]=-\sqrt{3}Q_{0}-Q'_{0}$ &
$\left[Q_{2},Q^{2}\right]=\sqrt{3}Q_{0}-Q'_{0}$ &
$\left[Q_{3},Q^{3}\right]=2Q'_{0}$ \\
$\left[Q_{1},Q_{2}\right]=\sqrt{2}Q^{3}$ &
$\left[Q_{2},Q_{3}\right]=\sqrt{2}Q^{1}$ &
$\left[Q_{3},Q_{1}\right]=\sqrt{2}Q^{2}$\\ \hline
\end{tabular}\\ \vspace{.2in}

The normalization in the above table is
$$Tr(Q_{0}Q_{0})=Tr(Q'_{0}Q'_{0})=Tr(Q_{1}Q^{1})=Tr(Q_{2}Q^{2})=Tr(Q_{3}Q^{3})=2$$

\begin{tabular}{|l|l|l|}\hline
\multicolumn{3}{|c|}{Table 8.1}\\  \multicolumn{3}{|c|}{
Non-trivial commutation relations of $E_{8}$ according to} \\
\multicolumn{3}{|c|}{the decomposition given by eq.(84)}\\ \hline
$\left[Q_{1},Q_{0}\right]=\sqrt{30}Q_{1}$ &
$\left[Q_{1},Q'_{0}\right]=\sqrt{10}Q_{1}$ &
$\left[Q_{2},Q_{0}\right]=-\sqrt{30}Q_{2}$ \\
$\left[Q_{2},Q'_{0}\right]=\sqrt{10}Q_{2}$ &
$\left[Q_{3},Q_{0}\right]=0$ &
$\left[Q_{3},Q'_{0}\right]=-2\sqrt{10}Q_{3}$ \\
$\left[Q_{1},Q^{1}\right]=-\sqrt{30}Q_{0}-\sqrt{10}Q'_{0}$ &
$\left[Q_{2},Q^{2}\right]=\sqrt{30}Q_{0}-\sqrt{10}Q'_{0}$ &
$\left[Q_{3},Q^{3}\right]=2\sqrt{10}Q'_{0}$ \\
$\left[Q_{1},Q_{2}\right]=\sqrt{20}Q^{3}$ &
$\left[Q_{2},Q_{3}\right]=\sqrt{20}Q^{1}$ &
$\left[Q_{3},Q_{1}\right]=\sqrt{20}Q^{2}$ \\
$\left[Q_{1i},Q_{0}\right]=\sqrt{30}Q_{1i}$ &
$\left[Q_{1i},Q'_{0}\right]=\sqrt{10}Q_{1i}$ &
$\left[Q_{2i},Q_{0}\right]=-\sqrt{30}Q_{2i}$ \\
$\left[Q_{2i},Q'_{0}\right]=\sqrt{10}Q_{2i}$ &
$\left[Q_{3i},Q_{0}\right]=0$ &
$\left[Q_{3i},Q'_{0}\right]=-2\sqrt{10}Q_{3i}$ \\
$\left[Q_{1i},Q_{2j}\right]=\sqrt{20}d_{ijk}Q^{3k}$ &
$\left[Q_{2i},Q_{3j}\right]=\sqrt{20}d_{ijk}Q^{1k}$ &
$\left[Q_{3i},Q_{1j}\right]=\sqrt{20}d_{ijk}Q^{2k}$ \\
$\left[Q^{\alpha},Q^{\beta}\right]=2ig^{\alpha\beta\gamma}Q^{\gamma}$
& $\left[Q^{\alpha},Q_{0}\right]=0$ &
$\left[Q^{\alpha},Q'_{0}\right]=0$ \\
$\left[Q^{\alpha},Q_{1i}\right]=-(G^{\alpha})^{j}_{i}Q_{1j}$ &
$\left[Q^{\alpha},Q_{2i}\right]=-(G^{\alpha})^{j}_{i}Q_{2j}$ &
$\left[Q^{\alpha},Q_{3i}\right]=-(G^{\alpha})^{j}_{i}Q_{3j}$\\\hline
\end{tabular} \\ \vspace{.2in}

\begin{tabular}{|c|}\hline
Table 8.2 \\ Further non-trivial commutation relations of
$E_{8}$\\ according to the decomposition given in eq.(84)\\ \hline
$\left[Q_{1i},Q^{1j}\right]=-\frac{1}{6}(G^{\alpha})^{j}_{i}Q^{\alpha}
-\sqrt{30}\delta^{j}_{i}Q_{0}-\sqrt{10}\delta^{j}_{i}Q'_{0}$ \\
$\left[Q_{2i},Q^{2j}\right]=-\frac{1}{6}(G^{\alpha})^{j}_{i}Q^{\alpha}
+\sqrt{30}\delta^{j}_{i}Q_{0}-\sqrt{10}\delta^{j}_{i}Q'_{0}$ \\
$\left[Q_{3i},Q^{3j}\right]=-\frac{1}{6}(G^{\alpha})^{j}_{i}Q^{\alpha}
+2\sqrt{10}\delta^{j}_{i}Q'_{0}$\\ \hline
\end{tabular} \\ \vspace{.2in}
The normalization is
$$Tr(Q_{0}Q_{0})=Tr(Q'_{0}Q'_{0})=Tr(Q_{1}Q^{1})=Tr(Q_{2}Q^{2})=Tr(Q_{3}Q^{3})=2.$$
$$Tr(Q_{1i}Q^{1j})=Tr(Q_{2i}Q^{2j})=Tr(Q_{3i}Q^{3j})=2\delta^{j}_{i}.$$
$$Tr(Q^{\alpha}Q^{\beta})=12\delta^{\alpha\beta}.$$
\end{centering}
\newpage
{\Large {\bf Appendix B.}} \\

Here we give some details related to the calculation of the V
operator in the case of $SU(3)/ U(1) \times U(1)$ and the gaugino
mass (92).

To calculate the V operator in the case of $SU(3)/U(1) \times
U(1)$ we use the real metric of the coset,
$g_{ab}=diag(a,a,b,b,c,c)$ with
$a=R_{1}^{2},b=R_{2}^{2},c=R_{3}^{2}$. Using the structure
constants of $SU(3)$, $ f_{12}^{\ \ 3}=2 $, $ f_{45}^{\ \
8}=f_{67}^{\ \ 8}=\sqrt{3} $, $ f_{24}^{\ \ 6} =f_{14}^{\ \
7}=f_{25}^{\ \ 7}=-f_{36}^{\ \ 7}=-f_{15}^{\ \ 6}=-f_{34}^{\ \
5}=1$, (where the indices 3 and 8 correspond to the $U(1) \times
U(1)$ and the rest are the coset indices) we calculate the
components of the $D_{abc}$:\\ $D_{523}
=D_{613}=D_{624}=D_{541}=-D_{514}=-D_{532}=-D_{631}=-D_{624}=\frac{1}{2}(c-a-b).$\\
$D_{235}=D_{136}=D_{624}=D_{154}=-D_{145}=-D_{253}=-D_{163}=-D_{264}=\frac{1}{2}(a-b-c).$\\
$D_{352}=D_{361}=D_{462}=D_{415}=-D_{451}=-D_{325}=-D_{316}=-D_{
426}=\frac{1}{2}(b-c-a).$\\ From the $D$'s we calculate the
contorsion tensor $$ \Sigma_{abc}=2\tau(D_{abc}+D_{bca}-D_{cba}),
$$ and then the tensor $$G_{abc}=D_{abc}+\frac{1}{2}\Sigma_{abc}$$
which is\\
$G_{523}=G_{613}=G_{642}=G_{541}=-G_{514}=-G_{532}=-G_{631}=-G_{642}=
\frac{1}{2}[(1-\tau)c-(1+\tau)a-(1+\tau)b].$\\
$G_{235}=G_{136}=G_{246}=G_{154}=-G_{145}=-G_{253}=-G_{163}=-G_{264}=
\frac{1}{2}[-(1-\tau)a+(1+\tau)b+(1+\tau)c].$\\
$G_{352}=G_{361}=G_{462}=G_{415}=-G_{451}=-G_{325}=-G_{316}=-G_{426}=
\frac{1}{2}[-(1+\tau)a+(1-\tau)b-(1+\tau)c].$\\ In addition we
need the gamma matrices. In ten dimensions we have $
\{\Gamma^{\mu},\Gamma^{\nu}\} = 2 \eta^{\mu\nu}$ with
$\Gamma^{\mu}= \gamma^{\mu}\otimes I_{8}$ and
$\{\Gamma^{a},\Gamma^{b}\} = -2g^{ab}$, where $$ \Gamma^{a} =
\frac{1}{\sqrt{r_{a}}}\gamma_{5}\otimes\left[\begin{array}{cc}0&\overline{\gamma}^{a}\\
\overline{\gamma}^{a}&0 \end{array} \right] $$ with $a=1,2,3,5,6$
and
$$\Gamma^{4}=\frac{1}{\sqrt{r_{4}}}\gamma_{5}\otimes\left[\begin{array}{cc}0&iI_{4}\\
iI_{4}&0 \end{array} \right]. $$ In the present case we have
$r_{1}=r_{2}=a$, $r_{3}=r_{4}=b$ and $r_{5}=r_{6}=c$. The
$\overline{\gamma}^{a}$ matrices are given by $
\overline{\gamma}^{1}=\sigma^{1}\otimes\sigma^{2}$ ,
$\overline{\gamma}^{2}=\sigma^{2}\otimes\sigma^{2}$,
$\overline{\gamma}^{3}=-I_{2}\otimes\sigma^{3}$,
$\overline{\gamma}^{5}=\sigma^{3}\otimes\sigma^{2}$,
$\overline{\gamma}^{6}=-I_{2}\otimes\sigma^{1}$. Using these
matrices we calculate
$\Sigma^{ab}=\frac{1}{4}[\Gamma^{a},\Gamma^{b}]$ and then
$G_{abc}\Gamma^{a}\Sigma^{bc}$.

\end{document}